\documentclass[aps,prc,twocolumn,showkeys,superscriptaddress,amsmath,amssymb,floatfix]{revtex4-1}

\usepackage[T1]{fontenc}
\usepackage{CJKutf8}
\usepackage{graphicx}
\usepackage{xcolor}
\usepackage{etoolbox}
\usepackage{threeparttable}
\usepackage{longtable}
\usepackage{upgreek}
\usepackage{amsmath}
\usepackage{amssymb}
\usepackage{lineno}
\usepackage[colorlinks=true,linkcolor=blue,citecolor=blue,urlcolor=blue]{hyperref}

\begin{document}
\begin{CJK*}{UTF8}{gbsn}

\title{A general decay form and absolute nuclear stability}
\thanks{Wen-Qiang Zhang was supported by the National Natural Science Foundation of China (Grant No. 12405140).}

\author{Wen-Qiang Zhang (张文强)}
\email{zwqzwq@impcas.ac.cn}
\affiliation{State Key Laboratory of Heavy Ion Science and Technology, Institute of Modern Physics, Chinese Academy of Sciences, Lanzhou 730000, China}
\affiliation{University of Chinese Academy of Sciences, Beijing 100049, China}

\author{Chong Qi (亓冲)}
\email{chongq@kth.se}
\affiliation{KTH, Alba Nova University Center, SE-10691 Stockholm, Sweden}

\begin{abstract}
A nucleus is conventionally called stable when no decay has been observed. We replace this phenomenological definition with a criterion based on conservation laws. For this purpose, we formulate a general decay form (GDF) for finite charge-neutral nuclear systems that includes all possible final configurations. A nucleus is absolutely stable if no allowed configuration has lower mass, assuming exact baryon- and lepton-number conservation. A global search over all nuclei with measured atomic masses yields 84 absolutely stable nuclei. Besides the criterion of absolute stability, we introduce a dimensionless measure $S_{\rm GDF}$ to characterize the partial half-lives of GDF-allowed decay channels at a coarse-grained level. We rank all naturally occurring isotopes according to their maximum $S_{\rm GDF}$ values, identify their favored decay modes, and estimate the corresponding partial half-lives. We further identify promising candidates for natural $\alpha$ and two-neutrino double-$\beta$ decays. The resulting framework also provides a nuclear-scale perspective on the far-future evolution of matter, in which nuclear decay drives matter toward combinations of these 84 absolutely stable nuclei before still slower processes of matter evolution. These findings extend the conventional phenomenological stable--radioactive classification to a conservation-law-based framework for nuclear stability and provide a systematic guide for future rare-decay searches.
\end{abstract}

\keywords{General decay form, absolute nuclear stability, rare nuclear decays}

\maketitle
\end{CJK*}

\section{\label{sec:Int}Introduction}
At the microscopic level, ordinary matter is built from atoms composed of the two lightest baryons, the proton and the neutron, and the lightest charged lepton, the electron. Which combinations of these constituents persist is governed by nuclear decay and depends on the timescale considered. The conventional notion of nuclear stability is largely phenomenological: nuclei for which no decay has so far been observed are commonly regarded as stable. This definition is adequate on present cosmological timescales, but becomes insufficient for understanding the decay-driven evolution of matter in the remote future of an open universe~\cite{Dyson1979,RevModPhys.69.337}.

At a more fundamental level, the strong interaction alone imposes no obvious upper limit on how many nucleons can be bound together. Nuclear matter, together with its extensions to hyperonic and quark matter at higher densities, indicates that arbitrarily large baryon number can in principle be confined by the strong interaction with no instability intrinsic to the binding itself~\cite{RevModPhys.89.015007}. The stability of finite nuclei instead reflects the interplay of the strong, weak, and electromagnetic interactions, which can drive nuclei toward configurations of lower mass. A nucleus is unstable not simply because the strong force fails to bind it, but because a lower-mass configuration exists and is reachable through radioactive decay. If no such energetically accessible pathway exists, even a weakly bound nucleus is stable against decay. The question of stability should then be posed in a more absolute sense: does any physically allowed and energetically lower final configuration exist? Nuclear stability should therefore be determined by whether such configurations are accessible under the interactions and exact conservation laws of nature.

Historically, the $\alpha$, $\beta$, and $\gamma$ decay modes have formed the cornerstones of our understanding of nuclear instability. In recent decades, individual basic decay modes have been extended to simultaneous multiparticle emission processes~\cite{PhysRevLett.59.2020,PhysRevC.8.216,PhysRevLett.53.1897,PhysRevLett.133.022502,PhysRevLett.110.152501,Blank2008,PFUTZNER2023104050,PhysRevLett.94.232501,PhysRevLett.127.012501,DENISOV2022137569,jxqh-6gpj}, including two-neutrino double-$\beta$ decay, whose half-lives typically exceed $10^{18}$ yr~\cite{PhysRevLett.59.2020}. Beyond multiparticle extensions of a single basic decay mode, different decay modes may also act in combination within a single decay process, as in simultaneous $\beta\gamma$ processes, including radiative $\beta$ decay~\cite{PhysRev.93.518,Nico2006} and virtual $\beta\gamma$ transitions~\cite{PhysRev.128.1776,PhysRevC.92.044305}, as well as in our recently proposed simultaneous $\alpha\beta$ decay~\cite{ZhangQi2026prl}. All of these decay modes obey the known exact conservation laws. By contrast, no nuclear decay mode that violates a known exact conservation law has been observed. Proton decay~\cite{NATH2007191} would require baryon-number violation, with experimental partial-lifetime lower limits reaching the $10^{34}$-yr scale~\cite{PhysRevD.102.112011}; neutrinoless double-$\beta$ decay~\cite{RevModPhys.80.481} would require lepton-number violation, with half-life limits reaching the $10^{26}$-yr level~\cite{jkf6-48j8}.

In this work, we adopt the following two assumptions: any energetically accessible decay channel consistent with the known exact conservation laws is allowed in principle, whereas any decay channel violating these laws is absolutely forbidden. On this basis, we formulate a general decay form (GDF) for finite nuclear systems that encompasses all possible final configurations, and use it to establish an absolute-stability criterion, yielding 84 absolutely stable nuclei. Beyond the binary classification, we introduce an entropy-like measure $S_{\rm GDF}$ to rank competing decay channels and estimate their partial half-lives. Applying it to naturally occurring isotopes, we identify their favored decay modes. These results also provide a nuclear-scale perspective on the far-future evolution of matter, identifying the absolutely stable nuclei toward which matter may evolve through decay alone before much slower collective processes become important.

\section{\label{sec:Met}Methodology}
For finite nuclear systems, we describe nuclear decay in terms of neutral atoms. Under the constraints of known conservation laws, the decay of a neutral parent atom $({}^{A}_{Z}\mathrm{X})_0$ can be expressed as
\begin{equation}
({}^{A}_{Z}\mathrm{X})_0
\rightarrow
\sum_i({}^{A_i}_{Z_i}\mathrm{Y}_i)_0
+N_n n
+r\bar{\nu}_e
+s\nu_e
+k(e^+e^-)
+N_{\gamma}\gamma .
\label{eq1}
\end{equation}
Equation~(\ref{eq1}) provides a general representation of all possible final configurations. We will call Eq.~(\ref{eq1}) the general decay form (GDF). Here, the subscript $0$ denotes neutral atoms used for mass bookkeeping and does not imply that the immediate decay products must be neutral. $N_n$, $r$, $s$, $k$, and $N_\gamma$ are the numbers of emitted neutrons, antineutrinos, neutrinos, electron--positron pairs, and photons, respectively. The GDF describes the final configuration as a whole, without specifying the fundamental interactions involved or requiring energetically allowed intermediate states. In this neutral-atom notation, electric-charge conservation is automatically incorporated, and accompanying electromagnetic processes such as internal conversion and subsequent atomic relaxation are not written explicitly but are included implicitly. The baryon-number and lepton-number conservation laws require
\begin{align} 
\sum_i A_i+N_n &= A , \label{eq2}\\ 
\sum_i Z_i-r+s &= Z . \label{eq3} 
\end{align}
In addition, energy conservation imposes the constraint
\begin{equation}
Q=\Bigl(
M[({}^{A}_{Z}\mathrm{X})_0]
-\sum_i M[({}^{A_i}_{Z_i}\mathrm{Y}_i)_0]
-N_n m_n-2k m_e
\Bigr)c^2>0 .
\label{eq4}
\end{equation}
Other particle species could in principle be included in Eq.~(\ref{eq1}) as long as no known conservation law is violated, but are not written explicitly here because their production thresholds are generally much higher than typical nuclear-decay $Q$ values.

To identify the lowest-mass ground-state configurations allowed by the GDF, it is sufficient to consider the first term on the right-hand side of Eq.~(\ref{eq1}). A free neutron cannot lower the final-state mass because it is heavier than a neutral hydrogen atom. Each emitted electron--positron pair increases the final-state mass by $2m_e$. For $p\to n$ conversions, electron capture (EC) can always be chosen instead of positron emission, so that the corresponding electron--positron pairs can be avoided. Thus, at fixed total nucleon number $A$, the ground-state configuration search reduces to determining the lowest possible final-state mass:
\begin{equation}
M_{\min}(A)=
\min_{\{A_i, Z_i\}:\,\sum_i A_i=A} \sum_i M[({}^{A_i}_{Z_i}\mathrm{Y}_i)_0].
\label{eq5}
\end{equation} 
Then, the maximum decay energy for a neutral atom $({}^{A}_{Z}\mathrm{X})_0$ is given by
\begin{equation}
Q_{\max}(A,Z)=\Bigl(M[({}^{A}_{Z}\mathrm{X})_0]-M_{\min}(A)\Bigr)c^2.
\label{eq6}
\end{equation}
In this sense, stability is not determined by binding in isolation; rather, it is a statement about whether an energetically accessible lower-mass configuration exists under the exact conservation laws in force. If the lowest-mass configuration coincides with the parent atom itself, then $Q_{\max}(A, Z)=0$. In this case, no energetically allowed decay channel exists within the GDF, and the nucleus is absolutely stable. Conversely, $Q_{\max}(A, Z)>0$ implies that a decay channel to a lower-mass final-state configuration exists within the GDF, and the nucleus is not absolutely stable under this criterion.

Beyond the criterion for absolute stability, we further construct a framework for ranking GDF-allowed decay channels of an unstable nucleus according to their estimated partial decay rates. We restrict the following calculations to one-step, non-photon decay channels, with the leptonic sector limited to charge-changing weak conversions. Within this restricted set, $k$ represents the number of $p\to n$ conversions realized through positron emission, and hence $0\le k\le s$. According to Fermi's golden rule, the partial decay width is determined by the available final-state phase space and the transition matrix elements. In a coarse-grained approximation, we assume that the dominant contributions to the partial decay width can be represented by three factors: a phase-space factor $\eta_Q$, a weak-conversion suppression factor $\eta_w$, and a fragment-formation and barrier-penetration factor $\eta_f$. We thus write the phenomenological partial decay width $\Gamma_{\rm GDF}$ as
\begin{equation}
\Gamma_{\rm GDF}\propto\eta_Q\eta_w\eta_f.
\label{eq7}
\end{equation}
Since these factors enter multiplicatively and span many orders of magnitude, we define a dimensionless entropy-like measure $S_{\rm GDF}$ as
\begin{equation}
S_{\rm GDF}=\ln\eta_Q+\ln\eta_w+\ln\eta_f.
\label{eq8}
\end{equation}
We call $S_{\rm GDF}$ entropy-like because $\ln\eta_Q$ is, up to a constant, the logarithm of the final-state phase-space volume, i.e., a Boltzmann-type entropy $\ln\Omega$ of the accessible final states, whereas $\ln\eta_w$ and $\ln\eta_f$ act as logarithmic penalties for reaching them. 

The phase-space factor is approximated as
\begin{equation}
\eta_Q=q^\xi ,
\label{eq9}
\end{equation}
where $q\equiv Q/(1~{\rm MeV})$ is the dimensionless decay energy. The exponent $\xi$ is taken as
\begin{equation}
\xi=
\frac{3N_{\rm frag}-5}{2}
+5(r+k)
+2(s-k)
+(r+s) .
\label{eq10}
\end{equation}
Here, $N_{\rm frag}$ is the total number of baryonic final fragments in Eq.~(\ref{eq1}), including free neutrons. The terms $5(r+k)$ and $2(s-k)$ represent the leading phase-space powers for $\beta^\mp$ and EC conversions, respectively. The last term, $r+s$, is a matching term that restores the standard weak-decay phase-space powers when the formula is applied to one-fragment weak decays. For example, Eq.~(\ref{eq10}) gives $\xi=5$ for allowed single-$\beta$ decay and $\xi=11$ for two-neutrino double-$\beta$ decay, consistent with their established leading phase-space scalings~\cite{Behrens1982,PhysRevLett.108.062502}. For bookkeeping, we also define the GDF order as
\begin{equation}
O_{\rm GDF}=N_{\rm frag}-1+r+s .
\label{eq11}
\end{equation}

The weak-conversion suppression factor $\eta_w$ is written as
\begin{equation}
\eta_w=\eta_W^{r+s},
\label{eq12}
\end{equation}
where $\eta_W$ denotes the decay-rate suppression associated with a single weak conversion. A dimensionless measure of the weak-interaction strength is given by $G_F E^2$, where $G_F$ is the Fermi constant with dimensions of $E^{-2}$ in natural units. Accordingly, the characteristic scale of $\eta_W$ is set by $(G_F E^2)^2$, which typically lies around the $10^{-20}$ scale for $E$ of a few MeV. We therefore adopt $\eta_W=10^{-20}$ as a benchmark value.

For decay channels with $N_{\rm frag}=1$, we set the fragment-related factor $\eta_f=1$. For channels with $N_{\rm frag}>1$, $\eta_f$ is written as
\begin{equation}
\eta_f
=
C_{\rm frag}
F_{\rm comb}
P_{\rm tun}.
\label{eq13}
\end{equation}
Here, $C_{\rm frag}$ is a phenomenological normalization constant, $F_{\rm comb}$ is a simplified combinatorial proxy for the fragment-preformation probability, and $P_{\rm tun}$ is the barrier penetrability associated with quantum tunneling. The factor $F_{\rm comb}$ is defined as the inverse combinatorial multiplicity of partitioning the parent nucleons into the specified fragments, without distinguishing protons from neutrons in the counting,
\begin{equation}
F_{\rm comb}
=
\frac{
\prod_{j=1}^{N_{\rm frag}} A_j!
\prod_{\kappa} n_{\kappa}!
}
{A!},
\label{eq14}
\end{equation}
where $A_j$ is the mass number of the $j$-th baryonic final fragment, and $n_{\kappa}$ denotes the multiplicity of identical fragments labeled by $\kappa$. The second product in the numerator removes the overcounting of indistinguishable identical fragments. For the barrier penetrability $P_{\rm tun}$, the present numerical implementation is restricted to binary fragment decompositions and includes only the Coulomb barrier for simplicity. Consequently, only channels with $N_{\rm frag}\le2$ are considered below when determining the maximum values of $S_{\rm GDF}$. Nevertheless, it is expected that the global maximum values of $S_{\rm GDF}$ are attained among these channels, since channels with three or more fragments are generally further suppressed by additional barriers and smaller values of $F_{\rm comb}$. $P_{\rm tun}$ is evaluated within the WKB approximation~\cite{PhysRev.113.1593} as
\begin{equation}
\ln P_{\rm tun}
=
-\frac{2}{\hbar c}
\int_{R_{\rm in}}^{R_{\rm out}}
\sqrt{
2\mu c^2\left[V_C(\rho)-Q\right]
}
\,d\rho .
\label{eq15}
\end{equation}
The Coulomb potential is $V_C(\rho)=Z_1Z_2e^2/\rho$~\cite{Qi2019}, where $\rho$ is the distance between the centers of the two fragments. The inner turning point is approximated by the contact distance $R_{\rm in}=r_0(A_1^{1/3}+A_2^{1/3})$, with $r_0=1.20$ fm, and the outer turning point is $R_{\rm out}=Z_1Z_2e^2/Q$. The reduced mass is $\mu=A_1A_2/(A_1+A_2)$ in atomic mass units (u). For channels involving weak conversions, $Q$ in Eq.~(\ref{eq15}) is approximated by the total decay energy, since the much stronger energy dependence of Coulomb tunneling compared with the charge-changing weak-conversion phase space generally favors energy partitions in which most of the decay energy goes into the relative motion of the nuclear fragments~\cite{ZhangQi2026prl}. Evaluating both $\eta_Q$ and $P_{\rm tun}$ at the full $Q$ overestimates the rates of such combined channels, since the energy carried by the leptons is not available for tunneling.  This effect is, however, small compared with the overall logarithmic lifetime scale of these channels. With $F_{\rm comb}$ and $P_{\rm tun}$ defined above, the normalization constant $C_{\rm frag}$ in Eq.~(\ref{eq13}) is introduced to place channels involving fragment emission only and those involving weak transitions only on a common scale defined by $S_{\rm GDF}$. Based on the empirical half-life systematics of representative $\alpha$ and $\beta$ decays at comparable decay energies~\cite{Sobhani2025}, and guided by the weak-transition benchmark $\eta_W=10^{-20}$, we estimate $C_{\rm frag}$ to be of order $10^{20}$ and take $C_{\rm frag}=10^{20}$ in the present calculations.

The framework above enables us to calculate $S_{\rm GDF}$ for the class of GDF-allowed decay channels considered here. By construction, a larger $S_{\rm GDF}$ corresponds to a higher partial decay rate, and vice versa. Thus, the decay channels of a given nucleus can be ranked according to their $S_{\rm GDF}$ values. From Eqs.~(\ref{eq7}) and (\ref{eq8}), we have $\Gamma_{\rm GDF}\propto e^{S_{\rm GDF}}$. The corresponding partial half-life $T_{1/2}^{\rm GDF}$ is then estimated as

\begin{equation}
\log\left[{T_{1/2}^{\rm GDF}}({\rm yr})\right]
=a-\frac{S_{\rm GDF}}{\ln 10},
\label{eq16}
\end{equation}
where the intercept $a$ is determined by fitting experimental half-life data. Given the simplifying assumptions and approximations involved in the present framework, the resulting values of $T_{1/2}^{\rm GDF}$ should be regarded as indicative, model-dependent estimates rather than precise predictions.

\begin{figure}[htb]
\centering
\includegraphics[width=0.36\textwidth]{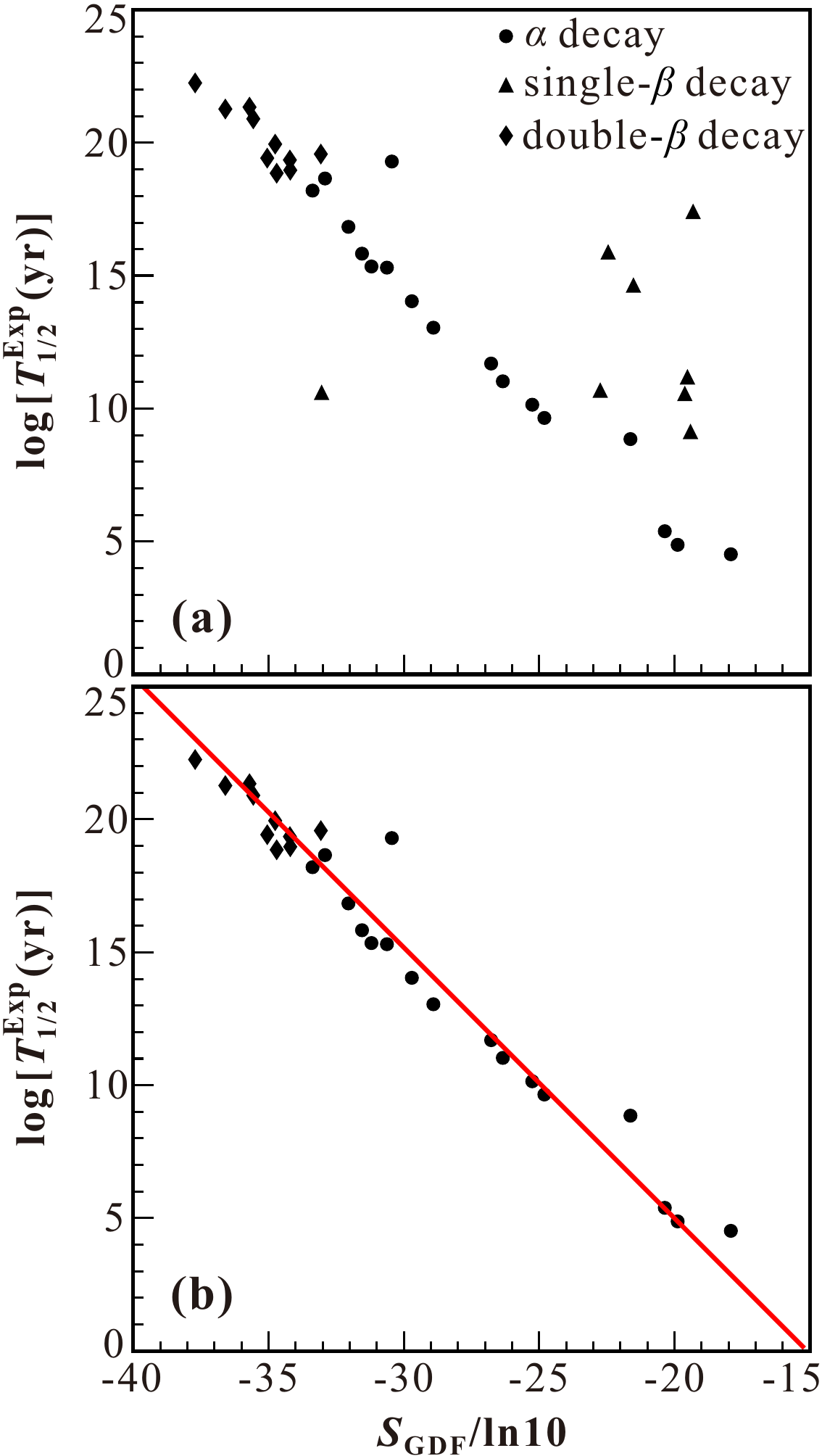}
\caption{Experimental half-lives as a function of $S_{\rm GDF}/\ln 10$ evaluated for the observed decay channels. (a) Results for 35 natural radioactive isotopes, including  $\alpha$,  single-$\beta$, and two-neutrino double-$\beta$ decays. (b) Same as (a), excluding single-$\beta$ decays. The solid line shows the fit using Eq.~(\ref{eq16}). Experimental data are taken from Ref.~\cite{NNDC}.}
\label{fig1}
\end{figure}

\begin{figure*}[htb]
\centering
\includegraphics[width=0.99\textwidth]{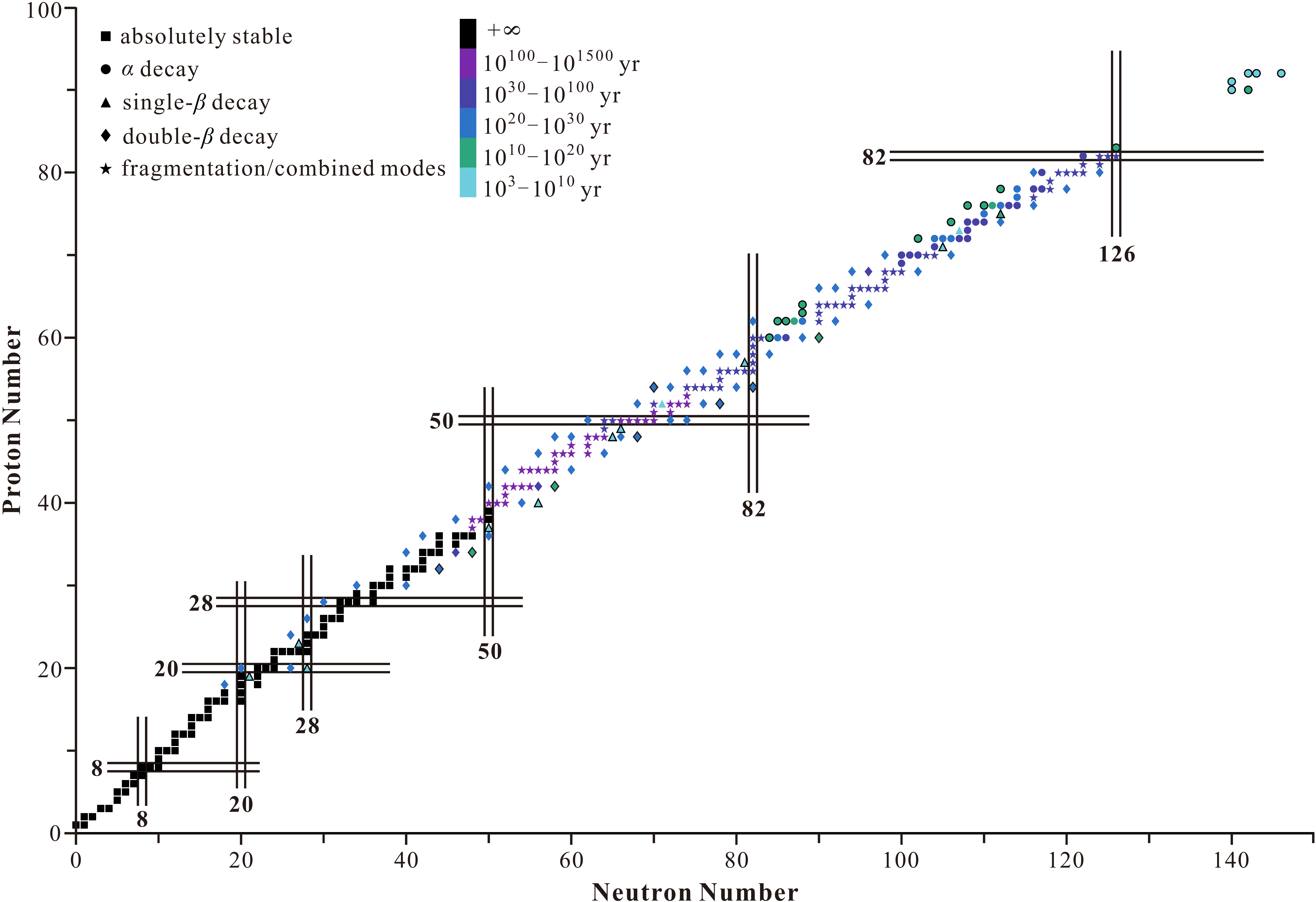}
\caption{GDF stability landscape of naturally occurring isotopes in the nuclear chart. Different dominant decay modes predicted by the GDF are represented by different symbols, with stars denoting fragmentation modes or combined modes involving multiple basic decay modes. Different colors indicate different ranges of the predicted $T_{1/2}^{\mathrm{GDF}}$. The 35 known natural radioactive isotopes are marked by black outlines. The positions of magic numbers are indicated by thick solid lines.}
\label{fig2}
\end{figure*}

\section{\label{sec:Dis}Results and Discussion}
We performed a global search for the lowest-mass ground-state configurations of finite charge-neutral nuclear systems with $1\le A\le300$ using available atomic mass data~\cite{CPC}. For each total nucleon number $A$, the minimum total mass $M_{\min}(A)$ was obtained from all possible neutral-fragment combinations, as defined in Eq.~(\ref{eq5}). The results identify 84 single-atom minima corresponding to absolutely stable nuclei under this criterion, with $A=1$--$4$, 6, 7, 9--84, 88, and 89, as listed in the first 84 entries of Table I in the Supplemental Material. For these nuclei, no energetically allowed decay channel exists, and we adopt the convention $S_{\rm GDF}^{\max}=-\infty$.

Beyond the stability classification, the GDF also provides a general framework for identifying possible decay modes of naturally occurring isotopes. Among the 289 naturally occurring isotopes~\cite{Holden2018}, radioactivity has so far been experimentally observed in 35 of them~\cite{NNDC}, leaving 254 that are presently regarded as stable. The GDF classification thus suggests that 170 of these observationally stable isotopes are potentially radioactive through energetically allowed decay channels. The possible decay modes of these nuclei have not previously been systematically explored. For each such nucleus, the channel with the maximum $S_{\rm GDF}$ value, $S_{\rm GDF}^{\max}$, is identified as the favored decay mode, with the corresponding GDF half-life $T_{1/2}^{\rm GDF}$ estimated from Eq.~(\ref{eq16}).

On this basis, we searched all naturally occurring isotopes for their $S_{\rm GDF}^{\max}$ values and corresponding dominant decay modes, and ranked them according to $S_{\rm GDF}^{\max}$, as listed in the Supplemental Material. Notably, all 35 experimentally known radioactive isotopes lie at the low-stability end of the ranking. Moreover, the GDF-favored decay modes agree with the experimentally observed modes for all of them except $^{48}$Ca and $^{96}$Zr, for which single-$\beta$ decay is predicted to dominate over the observed double-$\beta$ decay. The single-$\beta$ transitions of these two nuclei require $\Delta J\ge4$ and have not been observed.

With $S_{\rm GDF}$ determined, the intercept $a$ in Eq.~(\ref{eq16}) is required to estimate the corresponding $T_{1/2}^{\rm GDF}$. To determine $a$, we compared the calculated $S_{\rm GDF}$ values for the experimentally observed decay modes with the experimental half-lives of the 35 naturally occurring radioactive isotopes, as shown in Fig.~\ref{fig1}(a). The single-$\beta$ data deviate from the trend followed by the $\alpha$- and two-neutrino double-$\beta$-decay data. This can be attributed to a selection effect: naturally occurring single-$\beta$ emitters represent a strongly selected subset, since their survival to the present day is largely due to the hindrance of highly forbidden transitions by angular-momentum and parity selection rules not considered in $S_{\rm GDF}$, with $\log ft>10$ for all cases~\cite{NNDC}. In other words, the selection acts on the property that ensures survival over geological times. For natural $\alpha$ and double-$\beta$ emitters, survival is ensured by the low $Q$ values that suppress Coulomb-barrier tunneling~\cite{Qi2019} and by the second-order nature of the weak transition~\cite{Suhonen1998}, respectively. Both effects are included explicitly in $S_{\rm GDF}$, and the remaining nuclear-structure hindrance amounts to at most a few orders of magnitude, as for $^{209}$Bi (see Supplemental Material). The survival of these nuclei therefore does not preferentially select those with large hindrance from effects missing in $S_{\rm GDF}$. For single-$\beta$ emitters, by contrast, the property that ensures survival is precisely the one that $S_{\rm GDF}$ omits. We therefore excluded the single-$\beta$ decays from the fit and obtained $a=-14.843$ by fitting the remaining $\alpha$- and two-neutrino double-$\beta$-decay data with Eq.~(\ref{eq16}), as shown in Fig.~\ref{fig1}(b). 

The remaining data exhibit a clear linear correlation between $S_{\rm GDF}$ and $\log[T_{1/2}^{\rm Exp}({\rm yr})]$. The Pearson correlation coefficients are $-0.968$ for the $\alpha$ emitters and $-0.852$ for the double-$\beta$ emitters, indicating that the correlation is present in both decay modes.  Since the relative normalization of the two decay modes is set by $C_{\rm frag}$ and $\eta_W$, which were fixed at their order-of-magnitude estimates in Sec.~\ref{sec:Met}, the fact that both decay modes follow the same line provides a consistency check of the factorized width in Eq.~(\ref{eq7}). For $\alpha$ decay, the observed correlation is expected because the WKB penetrability in Eq.~(\ref{eq15}) reproduces the leading Coulomb-barrier tunneling dependence underlying Geiger-Nuttall-like laws~\cite{Qi2019,Geiger1911}. The $S_{\rm GDF}$ measure incorporates this well-established behavior and provides a unified basis for ranking competing channels across different decay modes.

As shown in Table I of the Supplemental Material, the estimated GDF half-lives extend into the $10^{100}$--$10^{1000}$ yr range. Although these estimates extend far beyond experimentally measured half-lives, they remain credible in terms of the overall logarithmic lifetime scale. This is because such ultralong lifetimes are primarily governed by decay-mode-intrinsic suppression largely independent of specific nuclear-structure hindrance. For example, the fragmentation decay $^{98}$Ru $\rightarrow$ $^{48}$Ti $+$ $^{50}$Ti, with $Q=11.7$ MeV, has a barrier penetrability of $P_{\rm tun}\approx10^{-158}$ calculated using Eq.~(\ref{eq15}), leading to an estimated GDF half-life of order $10^{150}$ yr. By comparison, even exceptionally large experimental $\alpha$-decay hindrance factors are only of order $10^7$~\cite{Kadmensky2007}, while $\beta$-decay $ft$ values span about 21 orders of magnitude~\cite{TURKAT2023101584}. Both are far smaller than the approximately 158 orders of magnitude of barrier suppression in this example. This suggests that nuclear-structure effects are unlikely to dominate the logarithmic scale of such ultralong-lived decays. Thus, despite not accounting for most detailed nuclear-structure effects, such as selection rules, level densities, shell structure, and deformation, $S_{\rm GDF}$ can still capture the dominant suppression mechanisms underlying these extreme lifetimes.

Figure~\ref{fig2} shows the overall systematics of the GDF-predicted decay modes and corresponding half-lives for naturally occurring isotopes. Three features are particularly evident: (1) all absolutely stable nuclei have neutron numbers that do not exceed the neutron magic number $N=50$; (2) $\alpha$ decay can become the dominant decay mode only beyond the neutron magic number $N=82$; and (3) nearly all isotopes with predicted $T_{1/2}^{\rm GDF}<10^{20}$ yr have been experimentally observed to be radioactive. The exceptions are the predicted $\alpha$ emitters $^{187}$Os and $^{149}$Sm discussed below and the EC decays of $^{123}$Te and $^{180m}$Ta. For the latter two, the GDF estimates of $10^{7.7}$ and $10^{5.1}$~yr lie far below the existing experimental lower limits~\cite{NNDC}. Both transitions are strongly forbidden, namely $1/2^+\to7/2^+$ for $^{123}$Te and a high-spin, $K$-forbidden transition from the $9^-$ isomer for $^{180m}$Ta, which is another manifestation of the selection effect discussed above. Taking $10^{30}$ yr as an optimistic limit for future experimental sensitivity~\cite{ZHANG2026}, 96 naturally occurring isotopes fall within this range, with their predicted dominant modes limited to $\alpha$, single-$\beta$, and double-$\beta$ decay. According to the predicted half-lives, the most promising candidates for future observation include $^{187}$Os, $^{149}$Sm, and $^{176,177}$Hf as new natural $\alpha$ emitters, and $^{124}$Sn, $^{110}$Pd, $^{148}$Nd, and $^{160}$Gd as new natural double-$\beta$ emitters. Owing to the aforementioned selection effect for naturally occurring single-$\beta$ emitters, we do not predict new natural single-$\beta$ candidates here. 

More broadly, the present results have implications for the far-future evolution of matter. If only nuclear decay is considered, decay-driven nuclear evolution must ultimately terminate in combinations of the 84 absolutely stable nuclei. Among the GDF-unstable nuclei, $^{87}$Sr is expected to have the longest half-life, of order $10^{1205}$ yr, while all others have substantially shorter predicted half-lives (see Supplemental Material). Nuclear decay would be essentially exhausted on a timescale of order $10^{1205}$ yr. However, this is still much shorter than the characteristic timescales for the subsequent fusion-driven and gravitational evolution of cold matter. Because $^{56}$Fe has the lowest atomic mass per nucleon~\cite{CPC}, even at zero temperature the absolutely stable nuclei with $A<56$ can continue to fuse toward iron through quantum tunneling, for which Dyson estimated a characteristic timescale of order $10^{1500}$ yr~\cite{Dyson1979}. This estimate refers to the complete conversion of matter to iron, which is limited by the reactions with the highest Coulomb barriers; fusion among the lightest nuclei may proceed considerably earlier. For absolutely stable nuclei with $A>56$, evolution toward $^{56}$Fe would require simultaneous fusion-fission processes and is therefore expected to occur on longer timescales. On vastly longer timescales, cold iron stars can further collapse into neutron stars or black holes, with characteristic timescales ranging from $10^{10^{26}}$ yr to $10^{10^{76}}$ yr depending on the underlying assumptions~\cite{Dyson1979}. Hence, there may exist an extraordinarily long intermediate epoch in an open universe, roughly $10^{1205}\ {\rm yr}\ll t\ll10^{1500}\ {\rm yr}$, during which decay-driven nuclear evolution has essentially ceased and matter is composed of combinations of the 84 absolutely stable nuclei, while the complete conversion to iron through fusion and the still slower gravitational evolution have not yet taken place. This scenario presupposes exact baryon-number conservation. If the proton decays, as predicted by grand unified theories, matter would disappear long before these timescales.

\section{\label{sec:Sum}Summary}
In summary, we have formulated a general decay form (GDF) for finite nuclear systems that encompasses all final configurations allowed by known conservation laws. A global search for the lowest-mass neutral configurations identifies 84 absolutely stable nuclei, for which no energetically allowed decay channel exists. For GDF-unstable nuclei, we introduced the dimensionless entropy-like measure $S_{\rm GDF}$ to rank competing decay channels and estimate their partial half-lives on an order-of-magnitude scale. Applying this framework to all naturally occurring isotopes, we find that 170 isotopes conventionally regarded as stable are energetically unstable within the GDF. In total, 96 naturally occurring isotopes have predicted half-lives below $10^{30}$ yr, taken here as an idealized sensitivity limit for rare-event searches, allowing us to identify several promising candidates for natural $\alpha$ and two-neutrino double-$\beta$ decays. The GDF thus places the conventional phenomenological stable--radioactive classification on a rigorous basis and offers a systematic guide for exploring rare and ultralong-lived decay processes. More broadly, on far-future timescales, these results suggest that decay-driven nuclear evolution would terminate in combinations of the 84 absolutely stable nuclei before the completion of the much slower fusion-driven and gravitational evolution of cold matter.

\bibliography{refGDF}

@article{DENISOV2022137569,
title = {Estimation of the double alpha-decay half-life},
journal = {Phys. Lett. B},
volume = {835},
pages = {137569},
year = {2022},
issn = {0370-2693},
doi = {https://doi.org/10.1016/j.physletb.2022.137569},
url = {https://www.sciencedirect.com/science/article/pii/S0370269322007031},
author = {V.Yu. Denisov}
}

@article{ZhangQi2026prl,
  title = {Simultaneous $\ensuremath{\alpha}\ensuremath{\beta}$ decay: A new mode of nuclear instability},
  author = {Zhang, W. Q. and Qi, C.},
  journal = {Phys. Rev. Lett.},
  year = {2026},
  month = {Nov},
  publisher = {American Physical Society},
  doi = {10.1103/76pc-tspc},
  url = {https://link.aps.org/doi/10.1103/76pc-tspc},
  note = {to be published}
}

@article{Geiger1911,
  author  = {H. Geiger and J. M. Nuttall},
  journal = {Philos. Mag.},
  volume  = {22},
  pages   = {613--621},
  year    = {1911},
  doi     = {10.1080/14786441008637156}
}

@article{PhysRev.128.1776,
  title = {Beta-Gamma Emission through Virtual States},
  author = {Rose, M. E. and Perrin, R. and Foldy, L. L.},
  journal = {Phys. Rev.},
  volume = {128},
  issue = {4},
  pages = {1776--1778},
  numpages = {0},
  year = {1962},
  month = {Nov},
  publisher = {American Physical Society},
  doi = {10.1103/PhysRev.128.1776},
  url = {https://link.aps.org/doi/10.1103/PhysRev.128.1776}
}

@article{PhysRevC.92.044305,
  title = {Evidence for the virtual $\ensuremath{\beta}\text{\ensuremath{-}}\ensuremath{\gamma}$ transition in $^{59}\mathrm{Ni}$ decay},
  author = {Pf{\"u}tzner, M. and Pachucki, K. and {\ifmmode \dot{Z}\else \.{Z}\fi{}}ylicz, J.},
  journal = {Phys. Rev. C},
  volume = {92},
  issue = {4},
  pages = {044305},
  numpages = {5},
  year = {2015},
  month = {Oct},
  publisher = {American Physical Society},
  doi = {10.1103/PhysRevC.92.044305},
  url = {https://link.aps.org/doi/10.1103/PhysRevC.92.044305}
}

@article{PhysRev.93.518,
  title = {Internal Bremsstrahlung and Ionization Accompanying Beta Decay},
  author = {Boehm, F. and Wu, C. S.},
  journal = {Phys. Rev.},
  volume = {93},
  issue = {3},
  pages = {518--523},
  numpages = {0},
  year = {1954},
  month = {Feb},
  publisher = {American Physical Society},
  doi = {10.1103/PhysRev.93.518},
  url = {https://link.aps.org/doi/10.1103/PhysRev.93.518}
}

@article{RevModPhys.69.337,
  title = {A dying universe: the long-term fate and evolution of astrophysical objects},
  author = {Adams, Fred C. and Laughlin, Gregory},
  journal = {Rev. Mod. Phys.},
  volume = {69},
  issue = {2},
  pages = {337--372},
  numpages = {0},
  year = {1997},
  month = {Apr},
  publisher = {American Physical Society},
  doi = {10.1103/RevModPhys.69.337},
  url = {https://link.aps.org/doi/10.1103/RevModPhys.69.337}
}

@article{RevModPhys.89.015007,
  title = {Equations of state for supernovae and compact stars},
  author = {Oertel, M. and Hempel, M. and Kl\"ahn, T. and others},
  journal = {Rev. Mod. Phys.},
  volume = {89},
  issue = {1},
  pages = {015007},
  numpages = {68},
  year = {2017},
  month = {Mar},
  publisher = {American Physical Society},
  doi = {10.1103/RevModPhys.89.015007},
  url = {https://link.aps.org/doi/10.1103/RevModPhys.89.015007}
}

@Article{Holden2018,
author={Holden, Norman and Coplen, Tyler and Bohlke, J. K. and others},
title={{IUPAC} Periodic Table of the Elements and Isotopes ({IPTEI}) for the education community ({IUPAC Technical Report})},
journal={Pure and Applied Chemistry},
year={2018},
volume={90},
number={12},
pages={1833-2092},
doi={10.1515/pac-2015-0703},
url={https://pubs.usgs.gov/publication/70208211},
url={https://doi.org/10.1515/pac-2015-0703}
}

@article{Dyson1979,
  title = {Time without end: Physics and biology in an open universe},
  author = {Dyson, Freeman J.},
  journal = {Rev. Mod. Phys.},
  volume = {51},
  issue = {3},
  pages = {447--460},
  numpages = {0},
  year = {1979},
  month = {Jul},
  publisher = {American Physical Society},
  doi = {10.1103/RevModPhys.51.447},
  url = {https://link.aps.org/doi/10.1103/RevModPhys.51.447}
}

@article{PhysRev.113.1593,
  title = {Alpha-Decay Barrier Penetrabilities with an Exponential Nuclear Potential: Even-Even Nuclei},
  author = {Rasmussen, John O.},
  journal = {Phys. Rev.},
  volume = {113},
  issue = {6},
  pages = {1593--1598},
  numpages = {0},
  year = {1959},
  month = {Mar},
  publisher = {American Physical Society},
  doi = {10.1103/PhysRev.113.1593},
  url = {https://link.aps.org/doi/10.1103/PhysRev.113.1593}
}

@Article{Sobhani2025,
author={Sobhani, Hadi
and Luo, Yan-An},
title={A unified formula for the half-life of the $\alpha$ and $\beta$ decay},
journal={Sci. Rep.},
year={2025},
month={Nov},
day={22},
volume={15},
number={1},
pages={41759},
issn={2045-2322},
doi={10.1038/s41598-025-23254-x},
url={https://doi.org/10.1038/s41598-025-23254-x}
}

@Article{Nico2006,
author={Nico, Jeffrey S. and Dewey, Maynard S. and Gentile, Thomas R. and others},
title={Observation of the radiative decay mode of the free neutron},
journal={Nature},
year={2006},
month={Dec},
day={01},
volume={444},
number={7122},
pages={1059-1062},
issn={1476-4687},
doi={10.1038/nature05390},
url={https://doi.org/10.1038/nature05390}
}

@article{PhysRevLett.59.2020,
  title = {Direct evidence for two-neutrino double-beta decay in $^{82}\mathrm{Se}$},
  author = {Elliott, S. R. and Hahn, A. A. and Moe, M. K.},
  journal = {Phys. Rev. Lett.},
  volume = {59},
  issue = {18},
  pages = {2020--2023},
  numpages = {0},
  year = {1987},
  month = {Nov},
  publisher = {American Physical Society},
  doi = {10.1103/PhysRevLett.59.2020},
  url = {https://link.aps.org/doi/10.1103/PhysRevLett.59.2020}
}

@article{PhysRevC.8.216,
  title = {Double Gamma Decay in $^{40}\mathrm{Ca}$},
  author = {Beardsworth, E. and Hensler, R. and Tape, J. W. and others},
  journal = {Phys. Rev. C},
  volume = {8},
  issue = {1},
  pages = {216--229},
  numpages = {0},
  year = {1973},
  month = {Jul},
  publisher = {American Physical Society},
  doi = {10.1103/PhysRevC.8.216},
  url = {https://link.aps.org/doi/10.1103/PhysRevC.8.216}
}

@article{PhysRevLett.127.012501,
  title = {Microscopic Description of $2\ensuremath{\alpha}$ Decay in $^{212}\mathrm{Po}$ and $^{224}\mathrm{Ra}$ Isotopes},
  author = {Mercier, F. and Zhao, J. and Ebran, J.-P. and others},
  journal = {Phys. Rev. Lett.},
  volume = {127},
  issue = {1},
  pages = {012501},
  numpages = {6},
  year = {2021},
  month = {Jul},
  publisher = {American Physical Society},
  doi = {10.1103/PhysRevLett.127.012501},
  url = {https://link.aps.org/doi/10.1103/PhysRevLett.127.012501}
}

@article{Blank2008,
doi = {10.1088/0034-4885/71/4/046301},
url = {https://doi.org/10.1088/0034-4885/71/4/046301},
year = {2008},
month = {mar},
publisher = {},
volume = {71},
number = {4},
pages = {046301},
author = {Blank, Bertram and Płoszajczak, Marek},
title = {Two-proton radioactivity},
journal = {Rep. Prog. Phys.}
}

@article{PhysRevLett.94.232501,
  title = {First Observation of $^{54}\mathrm{Zn}$ and its Decay by Two-Proton Emission},
  author = {Blank, B. and Bey, A. and Canchel, G. and others},
  journal = {Phys. Rev. Lett.},
  volume = {94},
  issue = {23},
  pages = {232501},
  numpages = {4},
  year = {2005},
  month = {Jun},
  publisher = {American Physical Society},
  doi = {10.1103/PhysRevLett.94.232501},
  url = {https://link.aps.org/doi/10.1103/PhysRevLett.94.232501}
}

@article{PhysRevLett.133.022502,
  title = {Measurement of the Isolated Nuclear Two-Photon Decay in $^{72}\mathrm{Ge}$},
  author = {Freire-Fern\'andez, D. and Korten, W. and Chen, R. J. and others},
  journal = {Phys. Rev. Lett.},
  volume = {133},
  issue = {2},
  pages = {022502},
  numpages = {7},
  year = {2024},
  month = {Jul},
  publisher = {American Physical Society},
  doi = {10.1103/PhysRevLett.133.022502},
  url = {https://link.aps.org/doi/10.1103/PhysRevLett.133.022502}
}

@book{Behrens1982,
  author    = {Behrens, H. and B{\"u}hring, W.},
  title     = {Electron Radial Wave Functions and Nuclear Beta-Decay},
  publisher = {Clarendon Press},
  address   = {Oxford},
  year      = {1982}
}

@Article{Kadmensky2007,
author={Kadmensky, S. G.
and Kurgalin, S. D.
and Tchuvil'sky, Yu. M.},
title={Cluster states in atomic nuclei and cluster-decay processes},
journal={Phys. Part. Nucl.},
year={2007},
month={Nov},
day={01},
volume={38},
number={6},
pages={699-742},
issn={1531-8559},
doi={10.1134/S1063779607060019},
url={https://doi.org/10.1134/S1063779607060019}
}

@article{TURKAT2023101584,
title = {Systematics of logft values for β−, and EC/β＋transitions},
journal = {At. Data Nucl. Data Tables},
volume = {152},
pages = {101584},
year = {2023},
issn = {0092-640X},
doi = {https://doi.org/10.1016/j.adt.2023.101584},
url = {https://www.sciencedirect.com/science/article/pii/S0092640X23000128},
author = {Steffen Turkat and Xavier Mougeot and Balraj Singh and others}
}

@article{PhysRevLett.108.062502,
  title = {$Q$ Value and Half-Lives for the Double-$\ensuremath{\beta}$-Decay Nuclide $^{110}\mathrm{Pd}$},
  author = {Fink, D. and Barea, J. and Beck, D. and others},
  journal = {Phys. Rev. Lett.},
  volume = {108},
  issue = {6},
  pages = {062502},
  numpages = {5},
  year = {2012},
  month = {Feb},
  publisher = {American Physical Society},
  doi = {10.1103/PhysRevLett.108.062502},
  url = {https://link.aps.org/doi/10.1103/PhysRevLett.108.062502}
}

@article{jkf6-48j8,
  author = {Abe, S. and others},
  collaboration = {KamLAND-Zen Collaboration},
  title = {Search for Majorana Neutrinos with the Complete KamLAND-Zen Dataset},
  journal = {Phys. Rev. Lett.},
  volume = {135},
  pages = {262501},
  year = {2025},
  doi = {10.1103/jkf6-48j8}
}

@article{Suhonen1998,
title = {Weak-interaction and nuclear-structure aspects of nuclear double beta decay},
journal = {Phys. Rep.},
volume = {300},
number = {3},
pages = {123-214},
year = {1998},
issn = {0370-1573},
doi = {https://doi.org/10.1016/S0370-1573(97)00087-2},
url = {https://www.sciencedirect.com/science/article/pii/S0370157397000872},
author = {Jouni Suhonen and Osvaldo Civitarese}
}

@article{ZHANG2026,
title = {Prediction for cluster radioactivity in even-even naturally occurring isotopes},
journal = {Phys. Lett. B},
volume = {878},
pages = {140559},
year = {2026},
issn = {0370-2693},
doi = {https://doi.org/10.1016/j.physletb.2026.140559},
url = {https://www.sciencedirect.com/science/article/pii/S0370269326004120},
author = {Zhang, W. Q. and Qi, C.}
}

@article{Qi2019,
title = {Recent developments in radioactive charged-particle emissions and related phenomena},
journal = {Prog. Part. Nucl. Phys.},
volume = {105},
pages = {214-251},
year = {2019},
issn = {0146-6410},
doi = {https://doi.org/10.1016/j.ppnp.2018.11.003},
url = {https://www.sciencedirect.com/science/article/pii/S0146641018301017},
author = {Chong Qi and Roberto Liotta and Ramon Wyss}
}

@article{PhysRevD.102.112011,
  title = {Search for proton decay via $p\ensuremath{\rightarrow}{e}^{+}{\ensuremath{\pi}}^{0}$ and $p\ensuremath{\rightarrow}{\ensuremath{\mu}}^{+}{\ensuremath{\pi}}^{0}$ with an enlarged fiducial volume in {Super-Kamiokande} {I}-{IV}},
  author = {{Super-Kamiokande Collaboration}},
  journal = {Phys. Rev. D},
  volume = {102},
  issue = {11},
  pages = {112011},
  year = {2020},
  month = {Dec},
  doi = {10.1103/PhysRevD.102.112011},
  url = {https://link.aps.org/doi/10.1103/PhysRevD.102.112011}
}

@misc{NNDC,
  author = {{National Nuclear Data Center}},
  title = {{NuDat 3.0 --- Nuclear Structure and Decay Data}},
  year = {2024},
  note = {{B}rookhaven National Laboratory},
  url = {https://www.nndc.bnl.gov/nudat3/},
}

@article{CPC,
title = {{The AME 2020 atomic mass evaluation (II). Tables, graphs and references}},
journal = {Chin. Phys. C},
volume = {45},
number = {3},
pages = {030003},
year = {2021},
doi = {10.1088/1674-1137/abddaf},
author = {{M. Wang} and {W. J. Huang} and {F. G. Kondev} and others},
}

@article{PFUTZNER2023104050,
title = {Two-proton emission and related phenomena},
journal = {Prog. Part. Nucl. Phys.},
volume = {132},
pages = {104050},
year = {2023},
issn = {0146-6410},
doi = {https://doi.org/10.1016/j.ppnp.2023.104050},
url = {https://www.sciencedirect.com/science/article/pii/S0146641023000315},
author = {M. Pfützner and I. Mukha and S.M. Wang}
}

@article{PhysRevLett.110.152501,
  title = {Study of Two-Neutron Radioactivity in the Decay of $^{26}\mathbf{O}$},
  author = {Kohley, Z. and Baumann, T. and Bazin, D. and others},
  journal = {Phys. Rev. Lett.},
  volume = {110},
  issue = {15},
  pages = {152501},
  numpages = {5},
  year = {2013},
  month = {Apr},
  publisher = {American Physical Society},
  doi = {10.1103/PhysRevLett.110.152501},
  url = {https://link.aps.org/doi/10.1103/PhysRevLett.110.152501}
}

@article{PhysRevLett.53.1897,
  title = {Double Gamma Decay in $^{40}\mathrm{Ca}$ and $^{90}\mathrm{Zr}$},
  author = {Schirmer, J. and Habs, D. and Kroth, R. and others},
  journal = {Phys. Rev. Lett.},
  volume = {53},
  issue = {20},
  pages = {1897--1900},
  numpages = {0},
  year = {1984},
  month = {Nov},
  publisher = {American Physical Society},
  doi = {10.1103/PhysRevLett.53.1897},
  url = {https://link.aps.org/doi/10.1103/PhysRevLett.53.1897}
}

@article{jxqh-6gpj,
  title = {Three-Body Barrier Dynamics of Double-Alpha Decay in Heavy Nuclei},
  author = {Tang, Shulin and Wan, Tao and Qian, Yibin and others},
  journal = {Phys. Rev. Lett.},
  volume = {136},
  issue = {17},
  pages = {172503},
  numpages = {6},
  year = {2026},
  month = {May},
  publisher = {American Physical Society},
  doi = {10.1103/jxqh-6gpj},
  url = {https://link.aps.org/doi/10.1103/jxqh-6gpj}
}

@article{RevModPhys.80.481,
  title = {Double beta decay, {M}ajorana neutrinos, and neutrino mass},
  author = {Avignone, Frank T. and Elliott, Steven R. and Engel, Jonathan},
  journal = {Rev. Mod. Phys.},
  volume = {80},
  issue = {2},
  pages = {481--516},
  numpages = {0},
  year = {2008},
  month = {Apr},
  publisher = {American Physical Society},
  doi = {10.1103/RevModPhys.80.481},
  url = {https://link.aps.org/doi/10.1103/RevModPhys.80.481}
}

@article{NATH2007191,
title = {Proton stability in grand unified theories, in strings and in branes},
journal = {Phys. Rep.},
volume = {441},
number = {5},
pages = {191-317},
year = {2007},
issn = {0370-1573},
doi = {https://doi.org/10.1016/j.physrep.2007.02.010},
url = {https://www.sciencedirect.com/science/article/pii/S0370157307000683},
author = {Pran Nath and Pavel {Fileviez Pérez}}
}

\onecolumngrid
\clearpage

\begin{center}
{\Large Supplemental Material}
\end{center}
\vspace{0.5em}

\setlength{\LTcapwidth}{\textwidth}
\providecommand{\nuc}[2]{}
\renewcommand{\nuc}[2]{\ensuremath{{}^{#1}\mathrm{#2}}}
\scriptsize
\renewcommand{\arraystretch}{1.2}
\setlength{\tabcolsep}{2.6pt}

\begin{longtable}{cccccccccc}
\caption{$S_{\mathrm{GDF}}$ ranking of naturally occurring isotopes, with all atomic species in the parent and final configurations treated as neutral atoms. The quantities $S_{\rm GDF}^{\max}$, $O_{\rm GDF}$, and $Q$ denote the maximum value of the entropy-like measure $S_{\rm GDF}$, the GDF order, and the decay energy, respectively. The GDF half-life estimate is obtained from $\log[T_{1/2}^{\mathrm{GDF}}(\mathrm{yr})]=a-S_{\mathrm{GDF}}^{\max}/\ln 10$, with $a=-14.843$. Absolute-stability entries are assigned the common rank 1. Experimental data are taken from Ref.~\cite{NNDC}. For $^{48}$Ca and $^{96}$Zr, the experimentally observed double-$\beta$ channels are additionally listed for comparison, with their channel-specific $S_{\rm GDF}$ values and estimated half-lives.}
\label{tab1}\\
\hline\hline
\noalign{\vskip 2pt}
Rank & Nuclide & $S_{\mathrm{GDF}}^{\max}$ & GDF-favored mode & Final state & $O_{\mathrm{GDF}}$ & $Q$~($\mathrm{MeV}$) & \shortstack{$\log[T_{1/2}^{\mathrm{GDF}}(\mathrm{yr})]$} & \shortstack{Exp. mode} & \shortstack{$\log[T_{1/2}^{\mathrm{Exp}}(\mathrm{yr})]$} \\
\noalign{\vskip 1pt}
\hline
\endfirsthead

\multicolumn{10}{l}{\textit{Table~\ref{tab1} continued.}}\\
\hline\hline
\noalign{\vskip 2pt}
Rank & Nuclide & $S_{\mathrm{GDF}}^{\max}$ & GDF-favored mode & Final state & $O_{\mathrm{GDF}}$ & $Q$~($\mathrm{MeV}$) & \shortstack{$\log[T_{1/2}^{\mathrm{GDF}}(\mathrm{yr})]$} & \shortstack{Exp. mode} & \shortstack{$\log[T_{1/2}^{\mathrm{Exp}}(\mathrm{yr})]$} \\
\noalign{\vskip 1pt}
\hline
\endhead

\hline
\multicolumn{10}{r}{\textit{Continued on next page}}\\
\endfoot

\hline\hline
\endlastfoot
1 & \nuc{1}{H} & $-\infty$ & - & - & - & - & $+\infty$ & - & - \\
1 & \nuc{2}{H} & $-\infty$ & - & - & - & - & $+\infty$ & - & - \\
1 & \nuc{3}{He} & $-\infty$ & - & - & - & - & $+\infty$ & - & - \\
1 & \nuc{4}{He} & $-\infty$ & - & - & - & - & $+\infty$ & - & - \\
1 & \nuc{6}{Li} & $-\infty$ & - & - & - & - & $+\infty$ & - & - \\
1 & \nuc{7}{Li} & $-\infty$ & - & - & - & - & $+\infty$ & - & - \\
1 & \nuc{9}{Be} & $-\infty$ & - & - & - & - & $+\infty$ & - & - \\
1 & \nuc{10}{B} & $-\infty$ & - & - & - & - & $+\infty$ & - & - \\
1 & \nuc{11}{B} & $-\infty$ & - & - & - & - & $+\infty$ & - & - \\
1 & \nuc{12}{C} & $-\infty$ & - & - & - & - & $+\infty$ & - & - \\
1 & \nuc{13}{C} & $-\infty$ & - & - & - & - & $+\infty$ & - & - \\
1 & \nuc{14}{N} & $-\infty$ & - & - & - & - & $+\infty$ & - & - \\
1 & \nuc{15}{N} & $-\infty$ & - & - & - & - & $+\infty$ & - & - \\
1 & \nuc{16}{O} & $-\infty$ & - & - & - & - & $+\infty$ & - & - \\
1 & \nuc{17}{O} & $-\infty$ & - & - & - & - & $+\infty$ & - & - \\
1 & \nuc{18}{O} & $-\infty$ & - & - & - & - & $+\infty$ & - & - \\
1 & \nuc{19}{F} & $-\infty$ & - & - & - & - & $+\infty$ & - & - \\
1 & \nuc{20}{Ne} & $-\infty$ & - & - & - & - & $+\infty$ & - & - \\
1 & \nuc{21}{Ne} & $-\infty$ & - & - & - & - & $+\infty$ & - & - \\
1 & \nuc{22}{Ne} & $-\infty$ & - & - & - & - & $+\infty$ & - & - \\
1 & \nuc{23}{Na} & $-\infty$ & - & - & - & - & $+\infty$ & - & - \\
1 & \nuc{24}{Mg} & $-\infty$ & - & - & - & - & $+\infty$ & - & - \\
1 & \nuc{25}{Mg} & $-\infty$ & - & - & - & - & $+\infty$ & - & - \\
1 & \nuc{26}{Mg} & $-\infty$ & - & - & - & - & $+\infty$ & - & - \\
1 & \nuc{27}{Al} & $-\infty$ & - & - & - & - & $+\infty$ & - & - \\
1 & \nuc{28}{Si} & $-\infty$ & - & - & - & - & $+\infty$ & - & - \\
1 & \nuc{29}{Si} & $-\infty$ & - & - & - & - & $+\infty$ & - & - \\
1 & \nuc{30}{Si} & $-\infty$ & - & - & - & - & $+\infty$ & - & - \\
1 & \nuc{31}{P} & $-\infty$ & - & - & - & - & $+\infty$ & - & - \\
1 & \nuc{32}{S} & $-\infty$ & - & - & - & - & $+\infty$ & - & - \\
1 & \nuc{33}{S} & $-\infty$ & - & - & - & - & $+\infty$ & - & - \\
1 & \nuc{34}{S} & $-\infty$ & - & - & - & - & $+\infty$ & - & - \\
1 & \nuc{35}{Cl} & $-\infty$ & - & - & - & - & $+\infty$ & - & - \\
1 & \nuc{36}{S} & $-\infty$ & - & - & - & - & $+\infty$ & - & - \\
1 & \nuc{37}{Cl} & $-\infty$ & - & - & - & - & $+\infty$ & - & - \\
1 & \nuc{38}{Ar} & $-\infty$ & - & - & - & - & $+\infty$ & - & - \\
1 & \nuc{39}{K} & $-\infty$ & - & - & - & - & $+\infty$ & - & - \\
1 & \nuc{40}{Ar} & $-\infty$ & - & - & - & - & $+\infty$ & - & - \\
1 & \nuc{41}{K} & $-\infty$ & - & - & - & - & $+\infty$ & - & - \\
1 & \nuc{42}{Ca} & $-\infty$ & - & - & - & - & $+\infty$ & - & - \\
1 & \nuc{43}{Ca} & $-\infty$ & - & - & - & - & $+\infty$ & - & - \\
1 & \nuc{44}{Ca} & $-\infty$ & - & - & - & - & $+\infty$ & - & - \\
1 & \nuc{45}{Sc} & $-\infty$ & - & - & - & - & $+\infty$ & - & - \\
1 & \nuc{46}{Ti} & $-\infty$ & - & - & - & - & $+\infty$ & - & - \\
1 & \nuc{47}{Ti} & $-\infty$ & - & - & - & - & $+\infty$ & - & - \\
1 & \nuc{48}{Ti} & $-\infty$ & - & - & - & - & $+\infty$ & - & - \\
1 & \nuc{49}{Ti} & $-\infty$ & - & - & - & - & $+\infty$ & - & - \\
1 & \nuc{50}{Ti} & $-\infty$ & - & - & - & - & $+\infty$ & - & - \\
1 & \nuc{51}{V} & $-\infty$ & - & - & - & - & $+\infty$ & - & - \\
1 & \nuc{52}{Cr} & $-\infty$ & - & - & - & - & $+\infty$ & - & - \\
1 & \nuc{53}{Cr} & $-\infty$ & - & - & - & - & $+\infty$ & - & - \\
1 & \nuc{54}{Cr} & $-\infty$ & - & - & - & - & $+\infty$ & - & - \\
1 & \nuc{55}{Mn} & $-\infty$ & - & - & - & - & $+\infty$ & - & - \\
1 & \nuc{56}{Fe} & $-\infty$ & - & - & - & - & $+\infty$ & - & - \\
1 & \nuc{57}{Fe} & $-\infty$ & - & - & - & - & $+\infty$ & - & - \\
1 & \nuc{58}{Fe} & $-\infty$ & - & - & - & - & $+\infty$ & - & - \\
1 & \nuc{59}{Co} & $-\infty$ & - & - & - & - & $+\infty$ & - & - \\
1 & \nuc{60}{Ni} & $-\infty$ & - & - & - & - & $+\infty$ & - & - \\
1 & \nuc{61}{Ni} & $-\infty$ & - & - & - & - & $+\infty$ & - & - \\
1 & \nuc{62}{Ni} & $-\infty$ & - & - & - & - & $+\infty$ & - & - \\
1 & \nuc{63}{Cu} & $-\infty$ & - & - & - & - & $+\infty$ & - & - \\
1 & \nuc{64}{Ni} & $-\infty$ & - & - & - & - & $+\infty$ & - & - \\
1 & \nuc{65}{Cu} & $-\infty$ & - & - & - & - & $+\infty$ & - & - \\
1 & \nuc{66}{Zn} & $-\infty$ & - & - & - & - & $+\infty$ & - & - \\
1 & \nuc{67}{Zn} & $-\infty$ & - & - & - & - & $+\infty$ & - & - \\
1 & \nuc{68}{Zn} & $-\infty$ & - & - & - & - & $+\infty$ & - & - \\
1 & \nuc{69}{Ga} & $-\infty$ & - & - & - & - & $+\infty$ & - & - \\
1 & \nuc{70}{Ge} & $-\infty$ & - & - & - & - & $+\infty$ & - & - \\
1 & \nuc{71}{Ga} & $-\infty$ & - & - & - & - & $+\infty$ & - & - \\
1 & \nuc{72}{Ge} & $-\infty$ & - & - & - & - & $+\infty$ & - & - \\
1 & \nuc{73}{Ge} & $-\infty$ & - & - & - & - & $+\infty$ & - & - \\
1 & \nuc{74}{Ge} & $-\infty$ & - & - & - & - & $+\infty$ & - & - \\
1 & \nuc{75}{As} & $-\infty$ & - & - & - & - & $+\infty$ & - & - \\
1 & \nuc{76}{Se} & $-\infty$ & - & - & - & - & $+\infty$ & - & - \\
1 & \nuc{77}{Se} & $-\infty$ & - & - & - & - & $+\infty$ & - & - \\
1 & \nuc{78}{Se} & $-\infty$ & - & - & - & - & $+\infty$ & - & - \\
1 & \nuc{79}{Br} & $-\infty$ & - & - & - & - & $+\infty$ & - & - \\
1 & \nuc{80}{Kr} & $-\infty$ & - & - & - & - & $+\infty$ & - & - \\
1 & \nuc{81}{Br} & $-\infty$ & - & - & - & - & $+\infty$ & - & - \\
1 & \nuc{82}{Kr} & $-\infty$ & - & - & - & - & $+\infty$ & - & - \\
1 & \nuc{83}{Kr} & $-\infty$ & - & - & - & - & $+\infty$ & - & - \\
1 & \nuc{84}{Kr} & $-\infty$ & - & - & - & - & $+\infty$ & - & - \\
1 & \nuc{88}{Sr} & $-\infty$ & - & - & - & - & $+\infty$ & - & - \\
1 & \nuc{89}{Y} & $-\infty$ & - & - & - & - & $+\infty$ & - & - \\
85 & \nuc{87}{Sr} & -2809.963 & 2frag+$2\beta^-$ & $\nuc{34}{S}+\nuc{53}{Cr}+2\bar{\nu}_e$ & 3 & 0.340 & 1205.508 &  & \\
86 & \nuc{85}{Rb} & -2632.891 & 2frag+$3\beta^-$ & $\nuc{29}{Si}+\nuc{56}{Fe}+3\bar{\nu}_e$ & 4 & 0.335 & 1128.607 &  & \\
87 & \nuc{86}{Sr} & -1726.609 & 2frag+$2\beta^-$ & $\nuc{34}{S}+\nuc{52}{Cr}+2\bar{\nu}_e$ & 3 & 0.829 & 735.014 &  & \\
88 & \nuc{90}{Zr} & -1235.287 & 2frag+$2\beta^-$ & $\nuc{34}{S}+\nuc{56}{Fe}+2\bar{\nu}_e$ & 3 & 1.766 & 521.635 &  & \\
89 & \nuc{91}{Zr} & -1082.613 & 2frag+$2\beta^-$ & $\nuc{34}{S}+\nuc{57}{Fe}+2\bar{\nu}_e$ & 3 & 2.218 & 455.330 &  & \\
90 & \nuc{92}{Zr} & -801.555 & 2frag+$2\beta^-$ & $\nuc{34}{S}+\nuc{58}{Fe}+2\bar{\nu}_e$ & 3 & 3.628 & 333.268 &  & \\
91 & \nuc{93}{Nb} & -683.978 & 2frag+$2\beta^-$ & $\nuc{34}{S}+\nuc{59}{Co}+2\bar{\nu}_e$ & 3 & 4.949 & 282.205 &  & \\
92 & \nuc{94}{Mo} & -608.511 & 2frag+$2\beta^-$ & $\nuc{38}{Ar}+\nuc{56}{Fe}+2\bar{\nu}_e$ & 3 & 6.908 & 249.430 &  & \\
93 & \nuc{95}{Mo} & -579.610 & 2frag+$\beta^-$ & $\nuc{37}{Cl}+\nuc{58}{Fe}+\bar{\nu}_e$ & 2 & 6.205 & 236.878 &  & \\
94 & \nuc{96}{Mo} & -519.474 & 2frag+$2\beta^-$ & $\nuc{34}{S}+\nuc{62}{Ni}+2\bar{\nu}_e$ & 3 & 7.883 & 210.762 &  & \\
95 & \nuc{97}{Mo} & -513.043 & 2frag+$\beta^-$ & $\nuc{47}{Sc}+\nuc{50}{Ti}+\bar{\nu}_e$ & 2 & 8.224 & 207.969 &  & \\
96 & \nuc{98}{Ru} & -380.777 & 2frag & $\nuc{48}{Ti}+\nuc{50}{Ti}$ & 1 & 11.700 & 150.526 &  & \\
97 & \nuc{99}{Ru} & -364.170 & 2frag & $\nuc{49}{Ti}+\nuc{50}{Ti}$ & 1 & 12.371 & 143.314 &  & \\
98 & \nuc{102}{Ru} & -354.043 & 2frag+$2\beta^-$ & $\nuc{50}{Ti}+\nuc{52}{Cr}+2\bar{\nu}_e$ & 3 & 17.746 & 138.916 &  & \\
99 & \nuc{101}{Ru} & -343.123 & 2frag+$\beta^-$ & $\nuc{50}{Ti}+\nuc{51}{V}+\bar{\nu}_e$ & 2 & 15.677 & 134.173 &  & \\
100 & \nuc{100}{Ru} & -333.947 & 2frag & $\nuc{50}{Ti}+\nuc{50}{Ti}$ & 1 & 13.637 & 130.188 &  & \\
101 & \nuc{106}{Pd} & -317.051 & 2frag & $\nuc{48}{Ca}+\nuc{58}{Fe}$ & 1 & 16.472 & 122.850 &  & \\
102 & \nuc{103}{Rh} & -311.003 & 2frag+$\beta^-$ & $\nuc{50}{Ti}+\nuc{53}{Cr}+\bar{\nu}_e$ & 2 & 18.688 & 120.224 &  & \\
103 & \nuc{108}{Pd} & -302.464 & 2frag+$2\beta^-$ & $\nuc{50}{Ti}+\nuc{58}{Fe}+2\bar{\nu}_e$ & 3 & 24.063 & 116.515 &  & \\
104 & \nuc{120}{Sn} & -293.493 & 2frag+$2\beta^-$ & $\nuc{32}{Si}+\nuc{88}{Sr}+2\bar{\nu}_e$ & 3 & 20.902 & 112.619 &  & \\
105 & \nuc{105}{Pd} & -292.429 & 2frag & $\nuc{51}{Ti}+\nuc{54}{Cr}$ & 1 & 18.250 & 112.157 &  & \\
106 & \nuc{118}{Sn} & -288.365 & 2frag+$2\beta^-$ & $\nuc{54}{Cr}+\nuc{64}{Ni}+2\bar{\nu}_e$ & 3 & 32.381 & 110.392 &  & \\
107 & \nuc{119}{Sn} & -287.027 & 2frag+$\beta^-$ & $\nuc{32}{Si}+\nuc{87}{Rb}+\bar{\nu}_e$ & 2 & 18.611 & 109.811 &  & \\
108 & \nuc{112}{Cd} & -285.675 & 2frag & $\nuc{48}{Ca}+\nuc{64}{Ni}$ & 1 & 20.749 & 109.224 &  & \\
109 & \nuc{109}{Ag} & -284.878 & 2frag+$\beta^-$ & $\nuc{50}{Ti}+\nuc{59}{Fe}+\bar{\nu}_e$ & 2 & 23.378 & 108.878 &  & \\
110 & \nuc{126}{Te} & -284.842 & 2frag+$2\beta^-$ & $\nuc{60}{Fe}+\nuc{66}{Ni}+2\bar{\nu}_e$ & 3 & 37.355 & 108.862 &  & \\
111 & \nuc{123}{Sb} & -283.158 & 2frag+$2\beta^-$ & $\nuc{36}{S}+\nuc{87}{Rb}+2\bar{\nu}_e$ & 3 & 26.039 & 108.131 &  & \\
112 & \nuc{111}{Cd} & -279.562 & 2frag & $\nuc{51}{Ti}+\nuc{60}{Fe}$ & 1 & 21.894 & 106.569 &  & \\
113 & \nuc{104}{Pd} & -279.432 & 2frag & $\nuc{50}{Ti}+\nuc{54}{Cr}$ & 1 & 18.972 & 106.513 &  & \\
114 & \nuc{107}{Ag} & -276.795 & 2frag & $\nuc{49}{Sc}+\nuc{58}{Fe}$ & 1 & 20.310 & 105.368 &  & \\
115 & \nuc{125}{Te} & -274.751 & 2frag+$2\beta^-$ & $\nuc{36}{S}+\nuc{89}{Sr}+2\bar{\nu}_e$ & 3 & 27.851 & 104.480 &  & \\
116 & \nuc{117}{Sn} & -272.807 & 2frag & $\nuc{51}{Ti}+\nuc{66}{Ni}$ & 1 & 25.341 & 103.636 &  & \\
117 & \nuc{124}{Te} & -272.351 & 2frag+$2\beta^-$ & $\nuc{36}{S}+\nuc{88}{Sr}+2\bar{\nu}_e$ & 3 & 28.062 & 103.438 &  & \\
118 & \nuc{110}{Cd} & -269.921 & 2frag & $\nuc{50}{Ti}+\nuc{60}{Fe}$ & 1 & 22.497 & 102.382 &  & \\
119 & \nuc{121}{Sb} & -268.760 & 2frag+$\beta^-$ & $\nuc{32}{Si}+\nuc{89}{Sr}+\bar{\nu}_e$ & 2 & 20.688 & 101.878 &  & \\
120 & \nuc{127}{I} & -266.359 & 2frag & $\nuc{48}{Ca}+\nuc{79}{As}$ & 1 & 28.878 & 100.835 &  & \\
121 & \nuc{116}{Sn} & -264.484 & 2frag & $\nuc{50}{Ti}+\nuc{66}{Ni}$ & 1 & 25.912 & 100.021 &  & \\
122 & \nuc{138}{Ba} & -263.875 & 2frag+$2\beta^-$ & $\nuc{48}{Ca}+\nuc{90}{Sr}+2\bar{\nu}_e$ & 3 & 41.914 & 99.756 &  & \\
123 & \nuc{113}{In} & -261.010 & 2frag & $\nuc{49}{Sc}+\nuc{64}{Ni}$ & 1 & 24.294 & 98.512 &  & \\
124 & \nuc{132}{Xe} & -260.897 & 2frag & $\nuc{48}{Ca}+\nuc{84}{Se}$ & 1 & 30.894 & 98.463 &  & \\
125 & \nuc{131}{Xe} & -257.825 & 2frag & $\nuc{48}{Ca}+\nuc{83}{Se}$ & 1 & 31.152 & 97.129 &  & \\
126 & \nuc{137}{Ba} & -257.262 & 2frag+$\beta^-$ & $\nuc{48}{Ca}+\nuc{89}{Rb}+\bar{\nu}_e$ & 2 & 38.216 & 96.884 &  & \\
127 & \nuc{115}{Sn} & -254.347 & 2frag & $\nuc{51}{Ti}+\nuc{64}{Ni}$ & 1 & 26.798 & 95.618 &  & \\
128 & \nuc{130}{Xe} & -249.730 & 2frag & $\nuc{48}{Ca}+\nuc{82}{Se}$ & 1 & 31.939 & 93.613 &  & \\
129 & \nuc{139}{La} & -249.717 & 2frag+$\beta^-$ & $\nuc{48}{Ca}+\nuc{91}{Sr}+\bar{\nu}_e$ & 2 & 40.655 & 93.608 &  & \\
130 & \nuc{129}{Xe} & -249.409 & 2frag & $\nuc{48}{Ca}+\nuc{81}{Se}$ & 1 & 31.918 & 93.474 &  & \\
131 & \nuc{136}{Ba} & -247.934 & 2frag & $\nuc{48}{Ca}+\nuc{88}{Kr}$ & 1 & 35.029 & 92.833 &  & \\
132 & \nuc{128}{Xe} & -246.952 & 2frag & $\nuc{48}{Ca}+\nuc{80}{Se}$ & 1 & 32.124 & 92.407 &  & \\
133 & \nuc{122}{Te} & -246.117 & 2frag & $\nuc{36}{S}+\nuc{86}{Kr}$ & 1 & 23.617 & 92.044 &  & \\
134 & \nuc{114}{Sn} & -240.549 & 2frag & $\nuc{50}{Ti}+\nuc{64}{Ni}$ & 1 & 27.971 & 89.626 &  & \\
135 & \nuc{140}{Ce} & -238.627 & 2frag & $\nuc{48}{Ca}+\nuc{92}{Sr}$ & 1 & 39.018 & 88.791 &  & \\
136 & \nuc{133}{Cs} & -237.821 & 2frag & $\nuc{48}{Ca}+\nuc{85}{Br}$ & 1 & 34.729 & 88.441 &  & \\
137 & \nuc{135}{Ba} & -230.402 & 2frag & $\nuc{48}{Ca}+\nuc{87}{Kr}$ & 1 & 37.084 & 85.219 &  & \\
138 & \nuc{141}{Pr} & -223.978 & 2frag & $\nuc{48}{Ca}+\nuc{93}{Y}$ & 1 & 42.437 & 82.429 &  & \\
139 & \nuc{134}{Ba} & -218.813 & 2frag & $\nuc{48}{Ca}+\nuc{86}{Kr}$ & 1 & 38.541 & 80.186 &  & \\
140 & \nuc{142}{Nd} & -208.179 & 2frag & $\nuc{56}{Cr}+\nuc{86}{Kr}$ & 1 & 52.601 & 75.568 &  & \\
141 & \nuc{143}{Nd} & -204.324 & $\alpha\mathrm{EC}$ & $\nuc{4}{He}+\nuc{139}{La}+\nu_e$ & 2 & 0.795 & 73.894 &  & \\
142 & \nuc{207}{Pb} & -189.963 & 2frag & $\nuc{79}{Ge}+\nuc{128}{Sn}$ & 1 & 130.439 & 67.657 &  & \\
143 & \nuc{205}{Tl} & -189.007 & 2frag & $\nuc{77}{Ga}+\nuc{128}{Sn}$ & 1 & 125.532 & 67.242 &  & \\
144 & \nuc{208}{Pb} & -188.855 & 2frag & $\nuc{80}{Ge}+\nuc{128}{Sn}$ & 1 & 131.147 & 67.176 &  & \\
145 & \nuc{206}{Pb} & -184.629 & 2frag & $\nuc{70}{Ni}+\nuc{136}{Xe}$ & 1 & 121.857 & 65.340 &  & \\
146 & \nuc{203}{Tl} & -183.116 & 2frag & $\nuc{71}{Cu}+\nuc{132}{Te}$ & 1 & 122.138 & 64.683 &  & \\
147 & \nuc{202}{Hg} & -182.850 & 2frag & $\nuc{70}{Ni}+\nuc{132}{Te}$ & 1 & 117.057 & 64.568 &  & \\
148 & \nuc{201}{Hg} & -180.642 & 2frag & $\nuc{69}{Ni}+\nuc{132}{Te}$ & 1 & 117.504 & 63.609 &  & \\
149 & \nuc{157}{Gd} & -178.343 & 2frag & $\nuc{32}{Si}+\nuc{125}{Sn}$ & 1 & 39.148 & 62.610 &  & \\
150 & \nuc{158}{Gd} & -176.803 & 2frag & $\nuc{32}{Si}+\nuc{126}{Sn}$ & 1 & 39.403 & 61.942 &  & \\
151 & \nuc{200}{Hg} & -175.163 & 2frag & $\nuc{68}{Ni}+\nuc{132}{Te}$ & 1 & 119.149 & 61.229 &  & \\
152 & \nuc{197}{Au} & -174.715 & 2frag & $\nuc{61}{Mn}+\nuc{136}{Xe}$ & 1 & 107.031 & 61.035 &  & \\
153 & \nuc{156}{Gd} & -174.366 & 2frag & $\nuc{32}{Si}+\nuc{124}{Sn}$ & 1 & 39.774 & 60.883 &  & \\
154 & \nuc{196}{Pt} & -173.668 & 2frag & $\nuc{68}{Ni}+\nuc{128}{Sn}$ & 1 & 114.180 & 60.580 &  & \\
155 & \nuc{199}{Hg} & -173.565 & 2frag & $\nuc{67}{Ni}+\nuc{132}{Te}$ & 1 & 119.385 & 60.535 &  & \\
156 & \nuc{159}{Tb} & -171.938 & 2frag & $\nuc{32}{Si}+\nuc{127}{Sb}$ & 1 & 41.243 & 59.829 &  & \\
157 & \nuc{195}{Pt} & -170.812 & $\alpha$ & $\nuc{4}{He}+\nuc{191}{Os}$ & 1 & 1.176 & 59.340 &  & \\
158 & \nuc{163}{Dy} & -168.578 & 2frag & $\nuc{32}{Si}+\nuc{131}{Te}$ & 1 & 42.908 & 58.369 &  & \\
159 & \nuc{162}{Dy} & -166.604 & 2frag & $\nuc{32}{Si}+\nuc{130}{Te}$ & 1 & 43.249 & 57.512 &  & \\
160 & \nuc{164}{Dy} & -166.425 & 2frag & $\nuc{32}{Si}+\nuc{132}{Te}$ & 1 & 43.298 & 57.434 &  & \\
161 & \nuc{173}{Yb} & -166.126 & 2frag & $\nuc{48}{Ca}+\nuc{125}{Sn}$ & 1 & 72.568 & 57.305 &  & \\
162 & \nuc{161}{Dy} & -165.978 & 2frag & $\nuc{26}{Mg}+\nuc{135}{Xe}$ & 1 & 34.572 & 57.240 &  & \\
163 & \nuc{160}{Dy} & -165.319 & 2frag & $\nuc{26}{Mg}+\nuc{134}{Xe}$ & 1 & 34.669 & 56.954 &  & \\
164 & \nuc{165}{Ho} & -164.541 & 2frag & $\nuc{28}{Mg}+\nuc{137}{Cs}$ & 1 & 36.668 & 56.616 &  & \\
165 & \nuc{174}{Yb} & -163.506 & 2frag & $\nuc{48}{Ca}+\nuc{126}{Sn}$ & 1 & 73.295 & 56.167 &  & \\
166 & \nuc{155}{Gd} & -157.524 & 2frag & $\nuc{16}{O}+\nuc{139}{Ba}$ & 1 & 17.581 & 53.569 &  & \\
167 & \nuc{167}{Er} & -157.322 & 2frag & $\nuc{32}{Si}+\nuc{135}{Xe}$ & 1 & 47.202 & 53.481 &  & \\
168 & \nuc{166}{Er} & -156.893 & 2frag & $\nuc{32}{Si}+\nuc{134}{Xe}$ & 1 & 47.280 & 53.295 &  & \\
169 & \nuc{168}{Er} & -155.771 & 2frag & $\nuc{32}{Si}+\nuc{136}{Xe}$ & 1 & 47.518 & 52.807 &  & \\
170 & \nuc{198}{Hg} & -155.244 & $\alpha$ & $\nuc{4}{He}+\nuc{194}{Pt}$ & 1 & 1.381 & 52.579 &  & \\
171 & \nuc{152}{Sm} & -153.425 & $\alpha+2\beta^-$ & $\nuc{4}{He}+\nuc{148}{Sm}+2\bar{\nu}_e$ & 3 & 2.148 & 51.789 &  & \\
172 & \nuc{153}{Eu} & -148.608 & 2frag & $\nuc{15}{N}+\nuc{138}{Ba}$ & 1 & 14.793 & 49.697 &  & \\
173 & \nuc{193}{Ir} & -147.469 & $\alpha\beta^-$ & $\nuc{4}{He}+\nuc{189}{Os}+\bar{\nu}_e$ & 2 & 2.026 & 49.202 &  & \\
174 & \nuc{190}{Os} & -144.541 & $\alpha$ & $\nuc{4}{He}+\nuc{186}{W}$ & 1 & 1.376 & 47.930 &  & \\
175 & \nuc{180}{Hf} & -141.468 & $\alpha$ & $\nuc{4}{He}+\nuc{176}{Yb}$ & 1 & 1.286 & 46.596 &  & \\
176 & \nuc{169}{Tm} & -141.215 & $\alpha$ & $\nuc{4}{He}+\nuc{165}{Ho}$ & 1 & 1.198 & 46.486 &  & \\
177 & \nuc{154}{Gd} & -140.455 & 2frag & $\nuc{16}{O}+\nuc{138}{Ba}$ & 1 & 19.293 & 46.156 &  & \\
178 & \nuc{194}{Pt} & -137.590 & $\alpha$ & $\nuc{4}{He}+\nuc{190}{Os}$ & 1 & 1.523 & 44.912 &  & \\
179 & \nuc{172}{Yb} & -133.692 & $\alpha$ & $\nuc{4}{He}+\nuc{168}{Er}$ & 1 & 1.309 & 43.219 &  & \\
180 & \nuc{181}{Ta} & -124.770 & $\alpha$ & $\nuc{4}{He}+\nuc{177}{Lu}$ & 1 & 1.520 & 39.344 &  & \\
181 & \nuc{184}{W} & -118.270 & $\alpha$ & $\nuc{4}{He}+\nuc{180}{Hf}$ & 1 & 1.649 & 36.521 &  & \\
182 & \nuc{204}{Pb} & -117.616 & $\alpha$ & $\nuc{4}{He}+\nuc{200}{Hg}$ & 1 & 1.968 & 36.237 &  & \\
183 & \nuc{183}{W} & -116.781 & $\alpha$ & $\nuc{4}{He}+\nuc{179}{Hf}$ & 1 & 1.672 & 35.874 &  & \\
184 & \nuc{98}{Mo} & -116.484 & $2\beta^-$ & $\nuc{98}{Ru}+2\bar{\nu}_e$ & 2 & 0.109 & 35.745 &  & \\
185 & \nuc{146}{Nd} & -115.706 & $\alpha$ & $\nuc{4}{He}+\nuc{142}{Ce}$ & 1 & 1.182 & 35.407 &  & \\
186 & \nuc{171}{Yb} & -114.436 & $\alpha$ & $\nuc{4}{He}+\nuc{167}{Er}$ & 1 & 1.557 & 34.856 &  & \\
187 & \nuc{80}{Se} & -114.212 & $2\beta^-$ & $\nuc{80}{Kr}+2\bar{\nu}_e$ & 2 & 0.134 & 34.759 &  & \\
188 & \nuc{175}{Lu} & -112.799 & $\alpha$ & $\nuc{4}{He}+\nuc{171}{Tm}$ & 1 & 1.619 & 34.145 &  & \\
189 & \nuc{182}{W} & -111.063 & $\alpha$ & $\nuc{4}{He}+\nuc{178}{Hf}$ & 1 & 1.764 & 33.391 &  & \\
190 & \nuc{164}{Er} & -110.548 & $2\mathrm{EC}$ & $\nuc{164}{Dy}+2\nu_e$ & 2 & 0.025 & 33.167 &  & \\
191 & \nuc{189}{Os} & -103.803 & $\alpha$ & $\nuc{4}{He}+\nuc{185}{W}$ & 1 & 1.976 & 30.238 &  & \\
192 & \nuc{179}{Hf} & -103.754 & $\alpha$ & $\nuc{4}{He}+\nuc{175}{Yb}$ & 1 & 1.808 & 30.217 &  & \\
193 & \nuc{170}{Yb} & -103.264 & $\alpha$ & $\nuc{4}{He}+\nuc{166}{Er}$ & 1 & 1.735 & 30.004 &  & \\
194 & \nuc{122}{Sn} & -102.951 & $2\beta^-$ & $\nuc{122}{Te}+2\bar{\nu}_e$ & 2 & 0.373 & 29.868 &  & \\
195 & \nuc{192}{Os} & -101.992 & $2\beta^-$ & $\nuc{192}{Pt}+2\bar{\nu}_e$ & 2 & 0.407 & 29.452 &  & \\
196 & \nuc{204}{Hg} & -101.646 & $2\beta^-$ & $\nuc{204}{Pb}+2\bar{\nu}_e$ & 2 & 0.420 & 29.301 &  & \\
197 & \nuc{150}{Sm} & -100.836 & $\alpha$ & $\nuc{4}{He}+\nuc{146}{Nd}$ & 1 & 1.450 & 28.950 &  & \\
198 & \nuc{191}{Ir} & -100.627 & $\alpha$ & $\nuc{4}{He}+\nuc{187}{Re}$ & 1 & 2.083 & 28.859 &  & \\
199 & \nuc{40}{Ca} & -100.303 & $2\mathrm{EC}$ & $\nuc{40}{Ar}+2\nu_e$ & 2 & 0.194 & 28.718 &  & \\
200 & \nuc{186}{W} & -99.928 & $2\beta^-$ & $\nuc{186}{Os}+2\bar{\nu}_e$ & 2 & 0.491 & 28.555 &  & \\
201 & \nuc{114}{Cd} & -98.780 & $2\beta^-$ & $\nuc{114}{Sn}+2\bar{\nu}_e$ & 2 & 0.545 & 28.057 &  & \\
202 & \nuc{108}{Cd} & -98.613 & $2\mathrm{EC}$ & $\nuc{108}{Pd}+2\nu_e$ & 2 & 0.272 & 27.984 &  & \\
203 & \nuc{158}{Dy} & -98.415 & $2\mathrm{EC}$ & $\nuc{158}{Gd}+2\nu_e$ & 2 & 0.283 & 27.898 &  & \\
204 & \nuc{170}{Er} & -96.741 & $2\beta^-$ & $\nuc{170}{Yb}+2\bar{\nu}_e$ & 2 & 0.656 & 27.171 &  & \\
205 & \nuc{36}{Ar} & -96.300 & $2\mathrm{EC}$ & $\nuc{36}{S}+2\nu_e$ & 2 & 0.432 & 26.980 &  & \\
206 & \nuc{188}{Os} & -95.631 & $\alpha$ & $\nuc{4}{He}+\nuc{184}{W}$ & 1 & 2.144 & 26.689 &  & \\
207 & \nuc{134}{Xe} & -94.233 & $2\beta^-$ & $\nuc{134}{Ba}+2\bar{\nu}_e$ & 2 & 0.824 & 26.082 &  & \\
208 & \nuc{54}{Fe} & -94.032 & $2\mathrm{EC}$ & $\nuc{54}{Cr}+2\nu_e$ & 2 & 0.680 & 25.995 &  & \\
209 & \nuc{138}{Ce} & -93.915 & $2\mathrm{EC}$ & $\nuc{138}{Ba}+2\nu_e$ & 2 & 0.696 & 25.944 &  & \\
210 & \nuc{128}{Te} & -93.673 & $2\beta^-$ & $\nuc{128}{Xe}+2\bar{\nu}_e$ & 2 & 0.867 & 25.839 &  & \\
211 & \nuc{196}{Hg} & -93.102 & $2\mathrm{EC}$ & $\nuc{196}{Pt}+2\nu_e$ & 2 & 0.819 & 25.591 &  & \\
212 & \nuc{132}{Ba} & -92.951 & $2\mathrm{EC}$ & $\nuc{132}{Xe}+2\nu_e$ & 2 & 0.844 & 25.525 &  & \\
213 & \nuc{126}{Xe} & -92.531 & $2\mathrm{EC}$ & $\nuc{126}{Te}+2\nu_e$ & 2 & 0.918 & 25.343 &  & \\
214 & \nuc{46}{Ca} & -92.236 & $2\beta^-$ & $\nuc{46}{Ti}+2\bar{\nu}_e$ & 2 & 0.988 & 25.215 &  & \\
215 & \nuc{70}{Zn} & -92.136 & $2\beta^-$ & $\nuc{70}{Ge}+2\bar{\nu}_e$ & 2 & 0.997 & 25.171 &  & \\
216 & \nuc{64}{Zn} & -91.650 & $2\mathrm{EC}$ & $\nuc{64}{Ni}+2\nu_e$ & 2 & 1.095 & 24.960 &  & \\
217 & \nuc{198}{Pt} & -91.567 & $2\beta^-$ & $\nuc{198}{Hg}+2\bar{\nu}_e$ & 2 & 1.050 & 24.924 &  & \\
218 & \nuc{50}{Cr} & -91.314 & $2\mathrm{EC}$ & $\nuc{50}{Ti}+2\nu_e$ & 2 & 1.171 & 24.814 &  & \\
219 & \nuc{185}{Re} & -91.270 & $\alpha$ & $\nuc{4}{He}+\nuc{181}{Ta}$ & 1 & 2.195 & 24.795 &  & \\
220 & \nuc{176}{Yb} & -91.206 & $2\beta^-$ & $\nuc{176}{Hf}+2\bar{\nu}_e$ & 2 & 1.085 & 24.767 &  & \\
221 & \nuc{102}{Pd} & -91.179 & $2\mathrm{EC}$ & $\nuc{102}{Ru}+2\nu_e$ & 2 & 1.203 & 24.756 &  & \\
222 & \nuc{74}{Se} & -91.154 & $2\mathrm{EC}$ & $\nuc{74}{Ge}+2\nu_e$ & 2 & 1.209 & 24.745 &  & \\
223 & \nuc{94}{Zr} & -90.614 & $2\beta^-$ & $\nuc{94}{Mo}+2\bar{\nu}_e$ & 2 & 1.145 & 24.510 &  & \\
224 & \nuc{168}{Yb} & -90.389 & $2\mathrm{EC}$ & $\nuc{168}{Er}+2\nu_e$ & 2 & 1.409 & 24.412 &  & \\
225 & \nuc{178}{Hf} & -89.834 & $\alpha$ & $\nuc{4}{He}+\nuc{174}{Yb}$ & 1 & 2.085 & 24.171 &  & \\
226 & \nuc{154}{Sm} & -89.649 & $2\beta^-$ & $\nuc{154}{Gd}+2\bar{\nu}_e$ & 2 & 1.250 & 24.091 &  & \\
227 & \nuc{92}{Mo} & -89.600 & $2\mathrm{EC}$ & $\nuc{92}{Zr}+2\nu_e$ & 2 & 1.650 & 24.070 &  & \\
228 & \nuc{86}{Kr} & -89.587 & $2\beta^-$ & $\nuc{86}{Sr}+2\bar{\nu}_e$ & 2 & 1.257 & 24.064 &  & \\
229 & \nuc{120}{Te} & -89.345 & $2\mathrm{EC}$ & $\nuc{120}{Sn}+2\nu_e$ & 2 & 1.736 & 23.959 &  & \\
230 & \nuc{104}{Ru} & -89.226 & $2\beta^-$ & $\nuc{104}{Pd}+2\bar{\nu}_e$ & 2 & 1.299 & 23.907 &  & \\
231 & \nuc{144}{Sm} & -89.215 & $2\mathrm{EC}$ & $\nuc{144}{Nd}+2\nu_e$ & 2 & 1.782 & 23.903 &  & \\
232 & \nuc{84}{Sr} & -89.195 & $2\mathrm{EC}$ & $\nuc{84}{Kr}+2\nu_e$ & 2 & 1.789 & 23.894 &  & \\
233 & \nuc{162}{Er} & -89.033 & $2\mathrm{EC}$ & $\nuc{162}{Dy}+2\nu_e$ & 2 & 1.848 & 23.824 &  & \\
234 & \nuc{112}{Sn} & -88.842 & $2\mathrm{EC}$ & $\nuc{112}{Cd}+2\nu_e$ & 2 & 1.920 & 23.741 &  & \\
235 & \nuc{58}{Ni} & -88.826 & $2\mathrm{EC}$ & $\nuc{58}{Fe}+2\nu_e$ & 2 & 1.926 & 23.734 &  & \\
236 & \nuc{156}{Dy} & -88.623 & $2\mathrm{EC}$ & $\nuc{156}{Gd}+2\nu_e$ & 2 & 2.006 & 23.645 &  & \\
237 & \nuc{142}{Ce} & -88.269 & $2\beta^-$ & $\nuc{142}{Nd}+2\bar{\nu}_e$ & 2 & 1.417 & 23.492 &  & \\
238 & \nuc{145}{Nd} & -88.003 & $\alpha$ & $\nuc{4}{He}+\nuc{141}{Ce}$ & 1 & 1.574 & 23.376 &  & \\
239 & \nuc{192}{Pt} & -87.896 & $\alpha$ & $\nuc{4}{He}+\nuc{188}{Os}$ & 1 & 2.423 & 23.330 &  & \\
240 & \nuc{136}{Ce} & -87.772 & $2\mathrm{EC}$ & $\nuc{136}{Ba}+2\nu_e$ & 2 & 2.378 & 23.276 &  & \\
241 & \nuc{130}{Ba} & -87.282 & $2\mathrm{EC}$ & $\nuc{130}{Xe}+2\nu_e$ & 2 & 2.623 & 23.063 &  & \\
242 & \nuc{96}{Ru} & -87.109 & $2\mathrm{EC}$ & $\nuc{96}{Mo}+2\nu_e$ & 2 & 2.715 & 22.988 &  & \\
243 & \nuc{106}{Cd} & -86.998 & $2\mathrm{EC}$ & $\nuc{106}{Pd}+2\nu_e$ & 2 & 2.776 & 22.940 &  & \\
244 & \nuc{78}{Kr} & -86.870 & $2\mathrm{EC}$ & $\nuc{78}{Se}+2\nu_e$ & 2 & 2.848 & 22.884 &  & \\
245 & \nuc{124}{Xe} & -86.855 & $2\mathrm{EC}$ & $\nuc{124}{Te}+2\nu_e$ & 2 & 2.857 & 22.878 & $2\mathrm{EC}$ & 22.255 \\
246 & \nuc{160}{Gd} & -86.074 & $2\beta^-$ & $\nuc{160}{Dy}+2\bar{\nu}_e$ & 2 & 1.730 & 22.538 &  & \\
247 & \nuc{148}{Nd} & -84.882 & $2\beta^-$ & $\nuc{148}{Sm}+2\bar{\nu}_e$ & 2 & 1.928 & 22.021 &  & \\
248 & \nuc{110}{Pd} & -84.386 & $2\beta^-$ & $\nuc{110}{Cd}+2\bar{\nu}_e$ & 2 & 2.017 & 21.805 &  & \\
249 & \nuc{76}{Ge} & -84.266 & $2\beta^-$ & $\nuc{76}{Se}+2\bar{\nu}_e$ & 2 & 2.039 & 21.753 & $2\beta^-$ & 21.276 \\
250 & \nuc{177}{Hf} & -83.017 & $\alpha$ & $\nuc{4}{He}+\nuc{173}{Yb}$ & 1 & 2.245 & 21.211 &  & \\
251 & \nuc{124}{Sn} & -82.975 & $2\beta^-$ & $\nuc{124}{Te}+2\bar{\nu}_e$ & 2 & 2.293 & 21.193 &  & \\
252 & \nuc{176}{Hf} & -82.676 & $\alpha$ & $\nuc{4}{He}+\nuc{172}{Yb}$ & 1 & 2.254 & 21.063 &  & \\
253 & \nuc{136}{Xe} & -82.211 & $2\beta^-$ & $\nuc{136}{Ba}+2\bar{\nu}_e$ & 2 & 2.458 & 20.861 & $2\beta^-$ & 21.342 \\
254 & \nuc{130}{Te} & -81.906 & $2\beta^-$ & $\nuc{130}{Xe}+2\bar{\nu}_e$ & 2 & 2.527 & 20.728 & $2\beta^-$ & 20.898 \\
255 & \nuc{116}{Cd} & -80.723 & $2\beta^-$ & $\nuc{116}{Sn}+2\bar{\nu}_e$ & 2 & 2.814 & 20.215 & $2\beta^-$ & 19.428 \\
256 & \nuc{82}{Se} & -80.026 & $2\beta^-$ & $\nuc{82}{Kr}+2\bar{\nu}_e$ & 2 & 2.998 & 19.912 & $2\beta^-$ & 19.944 \\
257 & \nuc{100}{Mo} & -79.895 & $2\beta^-$ & $\nuc{100}{Ru}+2\bar{\nu}_e$ & 2 & 3.034 & 19.855 & $2\beta^-$ & 18.849 \\
258 & \nuc{150}{Nd} & -78.736 & $2\beta^-$ & $\nuc{150}{Sm}+2\bar{\nu}_e$ & 2 & 3.371 & 19.352 & $2\beta^-$ & 18.968 \\
259 & \nuc{149}{Sm} & -77.665 & $\alpha$ & $\nuc{4}{He}+\nuc{145}{Nd}$ & 1 & 1.871 & 18.886 &  & \\
260 & \nuc{180}{W} & -76.840 & $\alpha$ & $\nuc{4}{He}+\nuc{176}{Hf}$ & 1 & 2.515 & 18.528 & $\alpha$ & 18.201 \\
261 & \nuc{187}{Re} & -76.068 & $\beta^-$ & $\nuc{187}{Os}+\bar{\nu}_e$ & 1 & 0.002 & 18.193 & $\beta^-$ & 10.615 \\
262 & \nuc{151}{Eu} & -75.787 & $\alpha$ & $\nuc{4}{He}+\nuc{147}{Pm}$ & 1 & 1.964 & 18.071 & $\alpha$ & 18.663 \\
263 & \nuc{174}{Hf} & -73.795 & $\alpha$ & $\nuc{4}{He}+\nuc{170}{Yb}$ & 1 & 2.494 & 17.206 & $\alpha$ & 16.845 \\
264 & \nuc{187}{Os} & -73.591 & $\alpha$ & $\nuc{4}{He}+\nuc{183}{W}$ & 1 & 2.722 & 17.117 &  & \\
265 & \nuc{148}{Sm} & -72.640 & $\alpha$ & $\nuc{4}{He}+\nuc{144}{Nd}$ & 1 & 1.987 & 16.704 & $\alpha$ & 15.833 \\
266 & \nuc{144}{Nd} & -71.848 & $\alpha$ & $\nuc{4}{He}+\nuc{140}{Ce}$ & 1 & 1.901 & 16.360 & $\alpha$ & 15.340 \\
267 & \nuc{186}{Os} & -70.523 & $\alpha$ & $\nuc{4}{He}+\nuc{182}{W}$ & 1 & 2.821 & 15.785 & $\alpha$ & 15.301 \\
268 & \nuc{209}{Bi} & -70.108 & $\alpha$ & $\nuc{4}{He}+\nuc{205}{Tl}$ & 1 & 3.137 & 15.605 & $\alpha$ & 19.303 \\
269 & \nuc{152}{Gd} & -68.400 & $\alpha$ & $\nuc{4}{He}+\nuc{148}{Sm}$ & 1 & 2.204 & 14.863 & $\alpha$ & 14.033 \\
270 & \nuc{184}{Os} & -66.575 & $\alpha$ & $\nuc{4}{He}+\nuc{180}{W}$ & 1 & 2.958 & 14.070 & $\alpha$ & 13.049 \\
271 & \nuc{190}{Pt} & -61.669 & $\alpha$ & $\nuc{4}{He}+\nuc{186}{Os}$ & 1 & 3.268 & 11.940 & $\alpha$ & 11.690 \\
272 & \nuc{147}{Sm} & -60.666 & $\alpha$ & $\nuc{4}{He}+\nuc{143}{Nd}$ & 1 & 2.311 & 11.504 & $\alpha$ & 11.029 \\
273 & \nuc{232}{Th} & -58.144 & $\alpha$ & $\nuc{4}{He}+\nuc{228}{Ra}$ & 1 & 4.082 & 10.409 & $\alpha$ & 10.148 \\
274 & \nuc{238}{U} & -57.126 & $\alpha$ & $\nuc{4}{He}+\nuc{234}{Th}$ & 1 & 4.270 & 9.967 & $\alpha$ & 9.650 \\
275 & \nuc{96}{Zr} & -55.091 & $\beta^-$ & $\nuc{96}{Nb}+\bar{\nu}_e$ & 1 & 0.164 & 9.083 &  & \\
275 & \nuc{96}{Zr} & -78.785 & $2\beta^-$ & $\nuc{96}{Mo}+2\bar{\nu}_e$ & 2 & 3.356 & 19.373 & $2\beta^-$ & 19.360 \\
276 & \nuc{48}{Ca} & -52.434 & $\beta^-$ & $\nuc{48}{Sc}+\bar{\nu}_e$ & 1 & 0.279 & 7.929 & $\beta^-$ & 20.120 \\
276 & \nuc{48}{Ca} & -76.141 & $2\beta^-$ & $\nuc{48}{Ti}+2\bar{\nu}_e$ & 2 & 4.268 & 18.225 & $2\beta^-$ & 19.570 \\
277 & \nuc{87}{Rb} & -52.381 & $\beta^-$ & $\nuc{87}{Sr}+\bar{\nu}_e$ & 1 & 0.282 & 7.906 & $\beta^-$ & 10.696 \\
278 & \nuc{123}{Te} & -51.965 & $\mathrm{EC}$ & $\nuc{123}{Sb}+\nu_e$ & 1 & 0.052 & 7.725 &  & \\
279 & \nuc{113}{Cd} & -51.687 & $\beta^-$ & $\nuc{113}{In}+\bar{\nu}_e$ & 1 & 0.324 & 7.604 & $\beta^-$ & 15.905 \\
280 & \nuc{235}{U} & -49.815 & $\alpha$ & $\nuc{4}{He}+\nuc{231}{Th}$ & 1 & 4.678 & 6.791 & $\alpha$ & 8.846 \\
281 & \nuc{115}{In} & -49.537 & $\beta^-$ & $\nuc{115}{Sn}+\bar{\nu}_e$ & 1 & 0.498 & 6.671 & $\beta^-$ & 14.644 \\
282 & \nuc{234}{U} & -46.886 & $\alpha$ & $\nuc{4}{He}+\nuc{230}{Th}$ & 1 & 4.858 & 5.519 & $\alpha$ & 5.390 \\
283 & \nuc{180m}{Ta} & -45.820 & $\mathrm{EC}$ & $\nuc{180}{Hf}+\nu_e$ & 1 & 1.123 & 5.056 &  & \\
284 & \nuc{230}{Th} & -45.786 & $\alpha$ & $\nuc{4}{He}+\nuc{226}{Ra}$ & 1 & 4.770 & 5.042 & $\alpha$ & 4.877 \\
285 & \nuc{176}{Lu} & -45.165 & $\beta^-$ & $\nuc{176}{Hf}+\bar{\nu}_e$ & 1 & 1.194 & 4.772 & $\beta^-$ & 10.570 \\
286 & \nuc{138}{La} & -44.934 & $\mathrm{EC}$ & $\nuc{138}{Ba}+\nu_e$ & 1 & 1.749 & 4.672 & $\beta^+/\mathrm{EC}$ & 11.196 \\
287 & \nuc{40}{K} & -44.702 & $\beta^-$ & $\nuc{40}{Ca}+\bar{\nu}_e$ & 1 & 1.310 & 4.571 & $\beta^-$ & 9.145 \\
288 & \nuc{50}{V} & -44.467 & $\mathrm{EC}$ & $\nuc{50}{Ti}+\nu_e$ & 1 & 2.209 & 4.469 & $\beta^+/\mathrm{EC}$ & 17.428 \\
289 & \nuc{231}{Pa} & -41.264 & $\alpha$ & $\nuc{4}{He}+\nuc{227}{Ac}$ & 1 & 5.150 & 3.078 & $\alpha$ & 4.515 \\

\end{longtable}

\normalsize

\end{document}